\documentclass[12pt,a4paper]{article}
\begin{document}

\title{Boundary Conformal Field Theories on Random Surfaces and The
Non-Critical Open String}
\author{Paul Mansfield and Rui Neves\thanks{Research suported by
J.N.I.C.T's PRAXIS XXI
PhD fellowship BD/2828/93-RM.}\\
\footnotesize{Department of Mathematical Sciences,
University of Durham, Science Laboratories}\\
\footnotesize{South Road,
Durham DH1 3LE, United Kingdom}\\
\footnotesize{P.R.W.Mansfield{\it @}Durham.ac.uk,
R.G.M.Neves{\it @}Durham.ac.uk}}
\maketitle

\begin{abstract}
We analyse boundary conformal field theories on random surfaces using
the conformal gauge approach of David, Distler and Kawai. The crucial
point is the choice of boundary conditions on the Liouville field. We
discuss the Weyl anomaly cancellation for Polyakov's
non-critical open bosonic string with Neumann, Dirichlet and free
boundary conditions. Dirichlet boundary conditions
on the Liouville
field imply that the metric is discontinuous as the boundary is
approached. We consider the semi-classical limit
and argue how it singles out the free boundary conditions for the
Liouville field. We define the open string susceptibility, the anomalous
gravitational scaling dimensions and a new Yang-Mills Feynman mass
critical exponent.
\end{abstract}

\section{Introduction}

In 1981 Polyakov \cite{Pol} showed that when non-critical strings are
quantised so as to maintain reparametrisation invariance, the scale of
the metric becomes a dynamical degree of freedom  even though it
decouples classically. Although the associated action is that of a
soluble quantum field theory, the Liouville theory, the integration
measure is not the usual one encountered in the functional approach to
quantum field theory. Consequently it was unclear how to proceed until
David, Distler and Kawai \cite{DDK} showed that the effect of the
measure could be accounted for by a simple renormalisation of the action.
In this paper we study the effects of boundaries on this approach,
extending their results to the cases of open string theory and to the
coupling of boundary conformal field theories to 2D quantum gravity.

We start in section 2 with a brief review of the coupling of the minimal
models to closed 2D quantum gravity.

In section 3 we consider our solution for the example of Polyakov's
non-critical open bosonic string. The key point is the choice of
boundary conditions on the Liouville field. Thus, we discuss the Weyl
anomaly cancellation for Neumann,
Dirichlet and free boundary conditions. We use a linear Coulomb gas
perturbative expansion \cite{MS,DF} to find the renormalised central
charge of the conformally
extended Liouville theory that describes the gravitational sector. As
expected
this will be shown to be the same central charge calculated for the
coupling on closed surfaces. Since the metric is to be written as a
reference metric multiplied by the exponential of the Liouville field
the theory must be independent of a shift in this field together with
a compensating Weyl transformation on the reference metric.
This leads to the dressing of
primary operators that acquire conformal weight $(1,1)$ on the bulk and
conformal weight $(1/2,1/2)$ on the boundary. Consequently, we show that
the Liouville field renormalisation is equal to the one found for closed
surfaces both on the bulk and on the boundary of the open surfaces. This
only works for Neumann and free boundary conditions on the Liouville
field. The Dirichlet boundary conditions freeze the Liouville boundary
quantum dynamics so that, it is not possible to cancel all the boundary
terms in the Weyl anomaly by a shift in the boundary values of the
Liouville field, without leading to a discontinuity in the metric as the
boundary is approached. Due to the presence of the boundary
we find new renormalised couplings to 2D gravity. Under Weyl
invariance at the quantum level we show that they are all determined by
the bulk or closed surface couplings as would be expected. We also show
how the Coulomb gas screening charge selection rule is a crucial
condition for the cancellation of non-local and Weyl anomalous
contributions to the correlation functions due to zero modes.

In section 4 we analyse the semi-classical limit
which singles out the free boundary conditions on the Liouville field
as being the most natural. We define the open string susceptibility,
the anomalous gravitational scaling dimensions and a new mass critical
exponent. In the context of Yang-Mills theory this mass exponent has an
interesting physical interpretation as the critical exponent associated
with the Feynman propagator for a test particle which interacts with the
gauge fields.

In section 5 we generalise the open string analysis to a
natural Feigin-Fuchs representation of $c\leq1$ minimal conformal field
theories on open random surfaces. Finally, we present our conclusions.

\section{Minimal Models On Closed Random Surfaces}

We now review the aspects of the approach of David, Distler
and Kawai \cite{DDK} to minimal models on closed random surfaces
that will be useful when we consider boundaries. The Coulomb gas
representation of conformal field theories due to
Dotsenko and Fateev \cite{MS,DF}  has a natural Lagrangian
interpretation. We introduce the action

\begin{eqnarray}
{S_M}[\Phi,\tilde{g}]&={1\over{8\pi}}\int{d^2}\xi\sqrt{\tilde{g}}
\left[{1\over{2}}{\tilde{g}^{ab}}{\partial_a}\Phi{\partial_b}
\Phi+i\left(\beta-1/\beta\right)\tilde{R}\Phi\right]+\nonumber\\
&+{\mu^2}\int{d^2}\xi\sqrt{\tilde{g}}\left({e^{i\beta\Phi}}+
{e^{-i/\beta\Phi}}\right)\label{30}
\end{eqnarray}

\noindent to define the minimal unitary series of conformal field
theories on closed surfaces. This is a conformally extended
Liouville theory \cite{PM} with imaginary coupling, $i\beta$,
on a surface with metric $\tilde{g}_{ab}$ and curvature $\tilde R$.
The central charge of the matter theory  is
${c_M}=1-6{{(\beta-1/\beta)}^2}$ which means the
minimal models \cite{BPZ} are at the rational points
${\beta^2}=(2+k')/(2+k)$. The
primary fields are vertex operators given by

\[
U(jj')=\int{d^2}\xi
\sqrt{\tilde{g}}\exp\left[-i\left(j\beta-{j'\over{\beta}}\right)
\Phi\right]
\]
where $j,j'\geq{0}$ are half-integer spins labelling pairs
of representations of the Virasoro algebra $A_1$. To couple this
theory to gravity we treat $\tilde{g}_{ab}$ as a dynamical
variable and add a cosmological constant term ${\mu_0^2}\int{d^2}\xi
\sqrt{\tilde{g}}$ to the action.

In the conformal gauge $\tilde{g}_{ab}$ is decomposed as a
reparametrisation of
${e^{\varphi}}{{\hat{g}}_{ab}}$. Integrating over the matter field and
reparametrisations generates a Weyl anomaly which yields a kinetic
term for $\varphi$ if the matter central charge
is not balanced by the corresponding reparametrisation ghost charge.
For this non-critical theory there results a Liouville field theory
for $\varphi$

\[
{S_L}\left[\varphi,\hat{g}\right]=-{{d-26}\over{48\pi}}\int{d^2}\xi
\sqrt{\hat{g}}\left({1\over{2}}\varphi\hat{\Delta}\varphi+
\hat{R}\varphi\right)+{\mu_1^2}\int{d^2}
\sqrt{\hat{g}}{e^{\varphi}},
\]
where $\hat\Delta$ is the covariant Laplacian
$-(1/\sqrt{\hat g}){\partial_a}\sqrt {\hat g}{{\hat g}^{ab}}
{\partial_b}$. The functional integral volume element for this theory
is induced by the inner product on variations of the Liouville field

\[
\left\|\delta\varphi\right\|^2_{\tilde{g}}=\int{d^2}\xi
\sqrt{\hat{g}}{e^{\varphi}}{{(\delta\varphi)}^2}.
\]
This theory is deeply
non-linear and its complete solution has not yet
been found \cite{DDK,LQT,GNM,KPZ}. The reason is the presence of
$e^{\varphi}$ in the inner product, which means that the
volume element is not the usual one that occurs in quantum field theory.
According to David, Distler and Kawai this may be replaced  by a
conventional field theory measure provided the Liouville mode and its
couplings to 2D quantum gravity are renormalised:

\[
{S_L}\left[\phi,\hat{g}\right]={1\over{8\pi}}\int{d^2}\xi
\sqrt{\hat{g}}\left[{1\over{2}}
\phi\hat{\Delta}\phi+i\left(\gamma+{1\over{\gamma}}\right)
\hat{R}\phi\right]+{\mu_2^2}\int{d^2}\sqrt{\hat{g}}{e^{\alpha\phi}}.
\]

Since the separation of ${\tilde g}_{ab}$ into the scale
$e^\phi$ and reference metric ${{\hat g}}_{ab}$ is arbitrary the new
theory is required to be invariant under simultaneous shifts in $\phi$
and compensating scalings of ${{\hat g}}_{ab}$. Thus a form of Weyl
invariance must be preserved at the quantum level. When we integrate
$\phi$ we generate a background Weyl anomaly which we add to the
background anomaly coming from the integration of the matter field and
the reparametrisation ghosts. The theory is Weyl invariant at the
quantum level if this anomaly is absent and the amplitude is independent
of the conformal factor of the reference metric.

The anomaly cancellation sets the total central charge of the system to
zero. This gives $\gamma=\pm{i}\beta$. Also the Liouville field
renormalisation parameter $\alpha$ must satisfy $1-\alpha(\beta+
1/\beta)+{\alpha^2}=0$ if we choose $\gamma=-i\beta$. Then we
have two branches ${\alpha_{+}}=\beta$ and ${\alpha_{-}}=1/\beta$. The
dressed vertex operators of vanishing conformal weight are

\[
{U_D}(jj')=\int{d^2}\xi\sqrt{\hat{g}}\exp\left[\left(l\beta-{l'\over
{\beta}}\right)\phi\right]\exp\left[-i\left(j\beta-{j'\over
{\beta}}\right)\Phi\right]
\]

\noindent where $l=-j$, $l'=j'+1$ or $l=j+1$, $l'=-j'$.

It is important to note that Weyl invariance at the quantum level is
only possible because we have imposed an independent charge
conservation selection rule \cite{DDK,MS,DF} on the matter and the
gravitational sectors.
For each sector the Gaussian integrals over $\Phi$ and $\phi$  yield
contributions of the form of the exponential of

\[
{\mathcal{F}^N}[g]=
{1\over{16\pi}}\int{d^2}\xi'{d^2}\xi''\sqrt{g(\xi')}{J^N}(\xi')
G(\xi',\xi'')
\sqrt{g(\xi'')}{J^N}(\xi'')
\]

\noindent where $g_{ab}$ stands for either $\tilde{g}_{ab}$ or
$\hat{g}_{ab}$, $J^N$ is the
coefficient of the term in the action that is linear in the field and
$G(\xi,\xi')$ is the
covariant Laplacian's Green's function which satisfies

\[
\Delta{G}(\xi,\xi')={{{\delta^2}(\xi-\xi')}\over{\sqrt{g(\xi)}}}-
{1\over{\int{d^2}\xi''
\sqrt{g(\xi'')}}},
\]
is symmetric in its arguments and orthogonal
to the constant zero-mode

\[
\int{d^2}\xi\sqrt{g(\xi)}G(\xi,\xi')=0.
\]

Due to the presence of the Laplacian's zero-mode we find a non-local
Weyl anomaly:

\begin{eqnarray*}
&{\delta_\rho}{\mathcal{F}^N}=-{Q\over{8\pi}}\int{d^2}\xi
\sqrt{g}{J^N}{\delta_\rho}\ln\int{d^2}\xi\sqrt{g}-
\nonumber\\
&-{1\over{8\pi\int{d^2}\xi\sqrt{g}}}\int{d^2}\xi
\sqrt{g}{J^N}\int{d^2}\xi'{d^2}\xi''
\sqrt{g(\xi')}\rho(\xi')G(\xi',\xi'')
\sqrt{g(\xi'')}{J^N}(\xi''),
\end{eqnarray*}

\noindent where $Q$ is either $i(\beta-1/\beta)$ or $i(\gamma+1/\gamma)$.
When we integrate the zero mode of the fields in each
sector the charge selection rule gives
$\int{d^2}\xi\sqrt{g}{J^N}=0$ for all non-zero contributions to the
amplitude, leading to the cancellation of the non-local anomaly.

Using a simple scaling argument David, Distler and Kawai's approach
leads us to the random surfaces critical exponents. We find the
susceptibility exponent \cite{DDK,ZCKT}
$\Gamma({\chi_c})=2-{\chi_c}(\beta+1/\beta)/
(2{\alpha_{\pm}})$, where ${\chi_c}=2-2h$ is the
Euler
characteristic of the closed Riemann surface given in terms of its
genus. It is related to the world-sheet integral of $\tilde{R}$ by the
Gauss-Bonnet theorem:

\begin{equation}
\int{d^2}\xi\sqrt{\tilde{g}}\tilde{R}=4\pi{\chi_c}\label{5}.
\end{equation}
The semi-classical limit
corresponds to $\beta\to+\infty$ and as expected it selects the
solution $\alpha_{+}=\beta$. We also get the gravitational scaling
dimensions of the matter primary fields \cite{DDK}
$\Delta(jj')=1-\beta(jj')_{\pm}/\alpha_{\pm}$. Here
$\beta(jj')_{\pm}$
defines the coefficient of the two possible dressings of the primary
field $U(jj')$. When this is combined with the bare conformal weight
using the equation which defines $\alpha$ it gives the
KPZ equation \cite{KPZ}:

\[
\Delta-{\Delta_0}=-{\alpha^2}\Delta(\Delta-1).
\]
These results for the critical exponents of a $c\leq1$
minimal conformal field theory on closed random surfaces agree with
the KPZ light-cone analysis on the sphere \cite{KPZ}. Distler, Hlousek
and Kawai \cite{DHK} also used this conformal gauge approach to calculate
the Hausdorff dimension of the random surfaces. All the results are in
striking agreement with those of the theory of dynamical triangulated
random surfaces \cite{KM}.

Our aim in this paper is to see if this picture still holds when the
random surfaces have boundaries and the minimal model becomes a boundary
conformal field theory. The main issue is the choice of boundary
conditions on the gravitational sector. Since the Liouville theory
has a natural generalisation in the presence of boundaries, we expect the
coupling of the minimal boundary conformal field theories to 2D quantum
gravity to be again described by two conformally extended Liouville
theories which are complementary. We start by presenting our solution in
the simple case of
Polyakov's open bosonic string.

\section{Open String 2D Quantum Gravity}

\subsection{Free boundary conditions}

For simplicity let us consider Polyakov's open bosonic string
partition function $Z$ for the topology of a disc \cite{Dur,OAL,Man}. We
take free
boundary conditions on the string field $X^{\mu}$ and on the Liouville
conformal gauge factor $\varphi$. In the case of the reparametrisation
ghosts $\theta^a$ we
consider diffeomorphisms which preserve the parameter domain in
$\mathcal{R}^2$ but allow for
general reparametrisations along the boundary. This means that the
component of $\theta^a$ along the outward normal to the boundary
must be zero $\tilde{n}\cdot\theta=0$, but its
component along the tangent $\tilde{t}\cdot\theta$ is kept free
just like the
boundary values of $X^{\mu}$ and $\varphi$. More precisely
we initially require that $X^{\mu}$,
$\varphi$ and $\tilde{t}\cdot\theta$ take
prescribed  values $Y^{\mu}$, $\psi$ and $\eta$ on the
boundary and then we integrate over these boundary values \cite{Man}.
The functional $Z[Y,\psi,\eta]$, obtained as an intermediate step, has
the physical interpretation of being the tree-level (in the sense of
string loops) contribution to the wave-functional of the vacuum for
closed string theory in the Schr\"odinger representation.

The quantum partition function
is thus given by

\[
Z=\int{\mathcal{D}_{\tilde{g}}}(Y,\psi,\eta)
Z[Y,\psi,\eta]
\]

\noindent where the wave functional is

\begin{equation}
Z[Y,\psi,\eta]=\int{\mathcal{D}_{\tilde{g}}}X
{\mathcal{D}_{\tilde{g}}}\tilde{g}\exp\left\{-
S[X,\tilde{g}]\right\}\label{29}.
\end{equation}

\noindent The action consists of the standard bosonic string
matter action of Brink, Di Vecchia and Howe plus renormalisation
counterterms:

\[
S[X,\tilde{g}]={1\over{16\pi}}\int{d^2}\xi\sqrt{\tilde{g}}
{\tilde{g}^{ab}}{\partial_{a}}{X^{\mu}}{\partial_{b}}{X^{\nu}}
{\eta_{\mu\nu}}+{\mu_0^2}
\int{d^2}\xi\sqrt{\tilde{g}}+{\lambda_0}\oint{d}\tilde{s}+{\nu_0}\oint
{d}\tilde{s}{k_{\tilde{g}}}.
\]

\noindent The cosmological constant
terms in the area ${\mu_0^2}\int{d^2}\xi\sqrt{\tilde{g}}$, the
invariant length of the boundary
${\lambda_0}\oint{d}\tilde{s}$ and the integral of its geodesic
curvature ${\nu_0}\oint{d}\tilde{s}{k_{\tilde{g}}}$
are the non-trivial pure gravity contributions to the action in two
dimensions. The first two are necessary as counterterms
due to short distance singularities. Although the geodesic curvature
counterterm is not associated with divergencies we will see that it is
absolutely necessary for our solution. Here we note that this term
can be written as $({\nu_0}/2)\int{d^2}\xi\sqrt{\tilde{g}}\tilde{R}$
if we use the Gauss-Bonnet theorem

\begin{equation}
\int{d^2}\xi\sqrt{\tilde{g}}\tilde{R}+2\oint{d}\tilde{s}
{k_{\tilde{g}}}=4\pi{\chi_o}\label{18}
\end{equation}

\noindent where ${\chi_o}$ is the Euler characteristic
of the
open Riemann surface. It is given by ${\chi_o}=2-2h-b$ where $h$
is the genus of the surface and $b$ the number of smooth boundaries.
Note also that in the open string the
Gauss-Bonnet theorem cannot fix both the integrals of the
scalar curvature $\tilde{R}$ and of the geodesic curvature
$k_{\tilde{g}}$ so that we should allow one of these as a pure
gravity contribution to the action.

To calculate $Z$ let us first determine the wave
functional $Z[Y,\psi,\eta]$. We start by separating $X^{\mu}$ into two
parts ${X^{\mu}}={X_c^{\mu}}+{\bar{X}^{\mu}}$. We define $X_c^{\mu}$
and $\bar{X}^{\mu}$ in such a way that the string action gets split
into two independent pieces, one for $X_c^{\mu}$ which contains all the
dependence on the boundary value $Y^{\mu}$ and
another for $\bar{X}^{\mu}$. This is easily done if we fix $X_c^{\mu}$
using $Y^{\mu}$,

\begin{equation}
\tilde{\Delta}{X_c^{\mu}}=0,\quad{X_c^{\mu}}{|_{B}}={Y^{\mu}},
\label{6}
\end{equation}

\noindent and impose on $\bar{X}^{\mu}$ a
homogeneous Dirichlet boundary
condition ${\bar{X}^{\mu}}{|_{B}}=0$. Here we have used the notation $B$
to say that the fields are evaluated at a point $\xi$ of the boundary $B$.
Eq. (\ref{6}) is solved in terms of $Y^{\mu}$ using
the homogeneous Green's function for the Laplacian with Dirichlet
boundary conditions defined for the
metric $\tilde{g}_{ab}$. We will separate the boundary value $Y^\mu$
into a constant piece, and a piece that is orthogonal with respect to
the natural metric on the boundary, i.e. we write
${Y^{\mu}}={Y_0^{\mu}}+{\bar{Y}^{\mu}}$ where
$\oint{d}\tilde{s}{\bar{Y}^{\mu}}=0$. Then if $\partial_{\tilde{n}}$
is the outward normal derivative on the boundary the solution is

\begin{equation}
{X_c^{\mu}}(\xi')={Y_0^{\mu}}-\oint{d}\tilde{s}(\xi)
{\partial_{\tilde{n}}}
{\tilde{G}_{D}}
(\xi,\xi'){\bar{Y}^{\mu}}(\xi)\label{7}
\end{equation}

\noindent if the point $\xi'$ is not in the boundary and
${X_c^{\mu}}{|_{B}}={Y^{\mu}}$ if it is. Of course here we have
considered

\begin{equation}
\tilde{\Delta}{\tilde{G}_D}(\xi,\xi')={{{\delta^2}(\xi-\xi')}\over
{\sqrt{\tilde{g}(\xi)}}}\label{8}
\end{equation}

\noindent where ${\tilde{G}_D}(\xi,\xi')=0$
if either argument lies on the boundary. In this case we can
integrate eq. (\ref{8}) leading to an integral
condition on its
outward normal derivative

\[
\oint{d}\tilde{s}(\xi){\partial_{\tilde{n}}}{\tilde{G}_D}(\xi,\xi')=-1,
\]
which allows the decomposition of $X_c^{\mu}$ given in
eq. (\ref{7}).

The string action can now be cast in
the form $S[X,\tilde{g}]={S_c}
[{X_c},\tilde{g}]+S[\bar{X},\tilde{g}]$. The action for
$\bar{X}^{\mu}$ is just the free bosonic action where the kinetic kernel
is the covariant
Laplacian. To find ${S_c}[{X_c},\tilde{g}]$ as a boundary action we take
a total derivative and use eq. (\ref{6}). We may write the result
introducing the boundary kinetic kernel
${\tilde{K}_D}(\xi,\xi')=-1/(8\pi)
{\partial_{\tilde{n}}}{\partial_{\tilde{n}'}}{\tilde{G}_D}
(\xi,\xi')$:

\[
{S_c}[{X_c},\tilde{g}]={1\over{2}}\oint{d}\tilde{s}(\xi)d\tilde{s}
(\xi')Y(\xi)\cdot{\tilde{K}_D}(\xi,\xi')Y(\xi').
\]

In standard fashion \cite{Pol,Pho,Dur,OAL,Man} the functional
integration measure ${\mathcal{D}_{\tilde{g}}}X$ is characterised
by an $\mathcal{L}^2$ norm for variations of $X^{\mu}$

\[
\left\|\delta{X}\right\|^2_{\tilde{g}}=\int{d^2}\xi\sqrt{\tilde{g}}
\delta
{X}\cdot\delta{X},\quad\int{\mathcal{D}_{\tilde{g}}}\delta{X}{e^{-
\left\|\delta{X}\right\|^2_{\tilde{g}}}}=1.
\]

\noindent When we integrate $X^{\mu}$ keeping $Y^{\mu}$ fixed
${\mathcal{D}_{\tilde{g}}}X$ is actually
${\mathcal{D}_{\tilde{g}}}\bar{X}$. For the integration over the metric
${\mathcal{D}_{\tilde{g}}}\tilde{g}$ we need to consider the similar
$\mathcal{L}^2$ norm for $\tilde{g}_{ab}$

\[
\left\|\delta\tilde{g}\right\|^2_{\tilde{g}}=\int{d^2}\xi
\sqrt{\tilde{g}}
\left({\tilde{g}^{ac}}{\tilde{g}^{bd}}+u{\tilde{g}^{ab}}
{\tilde{g}^{cd}}
\right)\delta{\tilde{g}_{ab}}\delta{\tilde{g}_{cd}}
\]

\noindent where $u$ is a non-negative constant. In the conformal gauge
we decompose
the integration over $\tilde{g}_{ab}$ into an integration over $\varphi$
and an integration over $\theta^a$. On the disc an arbitrary
infinitesimal variation of $\tilde{g}_{ab}$ is
$\delta{\tilde{g}_{ab}}=\delta\varphi
{\tilde{g}_{ab}}+{\tilde{\nabla}_a}
\delta{\theta_b}+{\tilde{\nabla}_b}\delta{\theta_a}$, where $\tilde
{\nabla}_a$ is the covariant derivative in the metric $\tilde{g}_{ab}$.
The variations of
$\tilde{g}_{ab}$ induced by the
reparametrisation ghosts and by Weyl transformations are not orthogonal.
They intersect in the conformal Killing vectors
${\tilde{P}_{ab}}(\delta\theta)=0$, where $\tilde{P}_{ab}$ acts on
vectors to make symmetric, traceless tensor fields
${\tilde{P}_{ab}}(\delta\theta)={\tilde{\nabla}_a}
\delta{\theta_b}+{\tilde{\nabla}_b}\delta{\theta_a}-{\tilde{g}_{ab}}
{\tilde{\nabla}_e}\delta{\theta^e}$. The adjoint acts on tensor fields
to make vectors ${\tilde{P}^{\dagger}_b}=-2{\tilde{\nabla}_a}{h^a_b}$.
Then redefining $\varphi$ we write

\[
\left\|\delta\tilde{g}\right\|^2_{\tilde{g}}=2(1+2u)\int{d^2}\xi
\sqrt{\tilde{g}}{{(\delta
\varphi)}^2}+\int{d^2}\xi\sqrt{\tilde{g}}{\tilde{g}^{ac}}
{\tilde{g}^{bd}}{\tilde{P}_{ab}}(\delta\theta){\tilde{P}_{cd}}(\delta
\theta).
\]

\noindent We now split $\theta^a$ into a field $\bar{\theta}^a$
vanishing at the boundary and another field $\vartheta^a$
such that at the boundary
${\vartheta^a}=\eta{\tilde{t}^a}$. Assuming that $\vartheta^a$ is fixed
by its boundary value in some way we obtain:

\[
\left\|\delta\tilde{g}\right\|^2_{\tilde{g}}=2(1+2u)\int{d^2}\xi
\sqrt{\tilde{g}}{{(\delta
\varphi)}^2}+\int{d^2}\xi\sqrt{\tilde{g}}\delta\bar{\theta}\cdot
{\tilde{P}^{\dagger}}\tilde{P}(\delta\bar{\theta}).
\]

Omiting the renormalisation counterterms we integrate $\bar{X}^{\mu}$
and $\bar{\theta}$ to find

\[
Z[Y,\psi,\eta]=\exp\left\{-
{S_c}[{X_c},\tilde{g}]\right\}\int{\mathcal{D}_{\tilde{g}}}\varphi
{{\left(Det'\tilde{\Delta}\right)}^{-d/2}}{{\sqrt{Det'
{\tilde{P}^{\dagger}}
\tilde{P}}}\over{Vol(CKV)}}
\]

\noindent where the prime denotes the omission of the zero modes and we
have divided by the volume of the space of conformal Killing vectors
$Vol(CKV)$. As
is well known these infinite determinants generate a Weyl
anomaly \cite{Pol,Pho,Dur,OAL,Man}. If we use the covariant heat kernel
to regularise them it is easy to see that the Weyl anomaly only depends
on the values of the heat
kernels for small proper time cutoff $\sqrt{\varepsilon}$. This means
that the Weyl anomaly is a local phenomenon which only reflects the
structure of the world-sheet at short distances. Since
$\sqrt{\varepsilon}$ can be made infinitesimally small, the bulk and
boundary contributions to the anomaly must be independent. Using
locality, reparametrisation invariance, dimensional analysis and the
commutativity of Weyl transformations we are led to the following
expansion in powers of the proper time cutoff $\sqrt{\varepsilon}$:

\begin{eqnarray*}
&{\delta_{\rho}}\ln\left[{{\left(Det'\tilde{\Delta}\right)}^{-d/2}}
{{\sqrt{Det'{\tilde{P}^{\dagger}}\tilde{P}}}\over{Vol(CKV)}}\right]=
{{d-26}\over{48\pi}}\int{d^2}\xi\sqrt{\tilde{g}}\tilde{R}\rho+
{{d-26}\over{24\pi}}\oint{d}\tilde{s}{k_{\tilde{g}}}\rho+\nonumber\\
&+{C_1}\oint{d}\tilde{s}{\partial_{\tilde{n}}}\rho+
{{C_2}\over{\varepsilon}}\int{d^2}\xi\sqrt{\tilde{g}}\rho+
{{C_3}\over{\sqrt{\varepsilon}}}\oint{d}\tilde{s}\rho+
O(\sqrt{\varepsilon}),
\end{eqnarray*}

\noindent where the $C_i$ are dimensionless constants which can be
determined exactly \cite{Dur,Man}. Here we will not worry about them
because all are
absorbed in the renormalisation counterterms.

Integrating the infinitesimal variation leads to the usual Liouville
action plus background contributions depending on the reference metric
of the conformal gauge $\hat{g}_{ab}$:

\[
Z[Y,\psi,\eta]=\exp\left\{-
{S_c}[{X_c},\tilde{g}]\right\}\int{\mathcal{D}_{\tilde{g}}}\varphi
{{\left(Det'\hat{\Delta}\right)}^{-d/2}}{{\sqrt{Det'{\hat{P}^
{\dagger}}
\hat{P}}}\over{Vol(CKV)}}\exp\left\{-
{S_L}[\varphi,\hat{g}]\right\},
\]

\noindent where the Liouville action is given by

\begin{eqnarray*}
{S_L}\left[\varphi,\hat{g}\right]&=-{{d-26}\over{48\pi}}\int{d^2}\xi
\sqrt{\hat{g}}\left({1\over{2}}{\hat{g}^{ab}}{\partial_a}
\varphi{\partial_b}\varphi+\hat{R}\varphi\right)-{{d-26}\over{24\pi}}
\oint{d}\hat{s}{k_{\hat{g}}}\varphi+\nonumber\\
&+{\mu_1^2}\int{d^2}
\sqrt{\hat{g}}{e^{\varphi}}+{\lambda_1}\oint{d}\hat{s}{e^{\varphi/2}}+
{\nu_1}\oint{d}\hat{s}{\partial_{\hat{n}}}\varphi.
\end{eqnarray*}

\noindent Here $\mu_1^2$, $\lambda_1$ and $\nu_1$ are arbitrary finite
constants left over from the renormalisation process.

Next we start the integration of the Liouville mode and determine the
renormalisation of the couplings to 2D quantum gravity.

\subsubsection{Anomaly cancellation for coupling
renormalisation}

To integrate the Liouville mode we start by taking the Coulomb gas
perturbative
approach expanding the area cosmological constant
counterterm. In each order of perturbation theory we split $\varphi$
in two fields $\varphi_c$, $\bar{\varphi}$ in exactly the same way we
split $X^{\mu}$ previously. As before the Liouville action becomes
the sum of two independent pieces, ${S_L}[{\varphi_c},\hat{g}]$, which
contains all the dependence on the boundary value
$\psi$, and ${S_L}[\bar{\varphi},\hat{g}]$. We further split
$\psi={\psi_0}+\bar{\psi}$
into a constant $\psi_0$ and an orthogonal piece
$\bar{\psi}$. The field $\varphi_c$ is now
expressed in terms of $\bar{\psi}$ and $\psi_0$:

\begin{equation}
{\varphi_c}(\xi')={\psi_0}-\oint{d}\hat{s}(\xi){\partial_{\hat{n}}}
{\hat{G}_D}
(\xi,\xi')\bar{\psi}(\xi)\label{13}.
\end{equation}

Let us take the lowest order in
the area cosmological constant perturbative expansion. When we
integrate $\varphi$ we consider a fixed value of
$\psi$. Then ${\mathcal{D}_{\tilde{g}}}\varphi=
{\mathcal{D}_{\tilde{g}}}\bar{\varphi}$ and the lowest order contribution
to the wave functional is given by

\[
{Z^{00}}[Y,\psi,\eta]=\exp\left\{-
{S_c}[{X_c},\tilde{g}]-{S_c^0}[{\varphi_c},\hat{g}]\right\}
{{\left(Det'\hat{\Delta}\right)}^{-d/2}}{{\sqrt{Det'{\hat{P}^
{\dagger}}
\hat{P}}}\over{Vol(CKV)}}{\bar{Z}^0}[Y,\psi,\eta]
\]

\noindent where

\[
{\bar{Z}^0}[Y,\psi,\eta]=\int{\mathcal{D}_{\tilde{g}}}\bar{\varphi}
\exp\left\{-{\bar{S}^0}[\bar{\varphi},\hat{g}]\right\}.
\]

\noindent Above we have introduced the lowest order Liouville actions
for $\bar{\varphi}$

\[
{\bar{S}^0}[\bar{\varphi},\hat{g}]=-{{d-26}\over{48\pi}}\int{d^2}
\xi\sqrt{\hat{g}}\left({1\over{2}}\bar{\varphi}\hat{\Delta}
\bar{\varphi}+\hat{R}\bar{\varphi}\right)+{\nu_1}\oint{d}\hat{s}
{\partial_{\hat{n}}}\bar{\varphi}
\]
and for $\varphi_c$

\begin{eqnarray}
&{S_c^0}\left[{\varphi_c},\hat{g}\right]=-{{d-26}\over{48\pi}}
\int{d^2}\xi
\sqrt{\hat{g}}\left({1\over{2}}{\hat{g}^{ab}}{\partial_a}
{\varphi_c}{\partial_b}{\varphi_c}+\hat{R}{\varphi_c}\right)-
\nonumber\\
&-{{d-26}\over{24\pi}}
\oint{d}\hat{s}{k_{\hat{g}}}{\varphi_c}+
{\lambda_1}\oint{d}\hat{s}{e^{{\varphi_c}/2}}\label{14}.
\end{eqnarray}

The functional integration measure for the integral over
$\bar{\varphi}$
is conformally invariant but non-linear in the Liouville field:

\[
\left\|\delta{\bar{\varphi}}\right\|^2_{\tilde{g}}=\int{d^2}\xi
\sqrt{\hat{g}}{e^{\varphi}}{{(\delta{\bar{\varphi}})}^2}.
\]

\noindent To proceed we need to use David, Distler and Kawai's
renormalisation ansatz \cite{DDK}. We may consider a canonical
measure in the background $\hat{g}_{ab}$,

\[
\left\|\delta{\bar{\phi}}\right\|^2_{\hat{g}}=\int{d^2}\xi
\sqrt{\hat{g}}{{(\delta{\bar{\phi}})}^2},
\]

\noindent provided we renormalise the Liouville field and its
couplings to 2D gravity. Observe that this renormalisation involves the
whole Liouville field. As pointed out by Symanzik in the presence of the
boundary we should expect to take independent
bulk and boundary renormalisations \cite{SYM}. Since the boundary pieces
of the Liouville mode are fixed at the moment we do not need to worry
about them for the time being. We also note that the canonical measure
can only be introduced if a set of background counterterms is included:

\[
{S_R}(\hat{g})={\mu_3^2}\int{d^2}\xi\sqrt{\hat{g}}+{\lambda_3}
\oint{d}\hat{s}+{\nu_3}\oint{d}\hat{s}{k_{\hat{g}}}.
\]

\noindent When we renormalise the field
$\bar{\varphi}\to\alpha\bar{\phi}$ and its couplings to gravity we get
the following renormalised lowest order Liouville action:

\[
{\bar{S}^0}[\bar{\phi},\hat{g}]={1\over{8\pi}}\int{d^2}
\xi\sqrt{\hat{g}}\left({1\over{2}}\bar{\phi}\hat{\Delta}\bar{\phi}+
Q\hat{R}\bar{\phi}\right)+{\nu_2}\oint{d}\hat{s}{\partial_{\hat{n}}}
\bar{\phi}.
\]

The renormalised parameters of the theory are determined by requiring
invariance under a shift in $\phi$ and a compensating Weyl
transformation of the reference metric. Once $\phi$ has been integrated
out the result is required to be invariant under Weyl transformations of
the metric alone. For the moment we integrate
$\bar{\phi}$. To do so we need to follow Alvarez \cite{OAL} and set
$\nu_2$ to zero because the standard way to deal with a term that is
linear in the field is to shift the integration variable, in this case
by a constant,
but this would spoil the homogeneous Dirichlet condition on $\bar\phi$.
Next we change variables as follows
$\sqrt{8\pi}\bar{\phi}\to\bar{\phi}+{\hat{O}_Q^0}$. Here we have set

\[
{\hat{O}_Q^0}
(\xi')=\int{d^2}\xi\sqrt{\hat{g}(\xi)}{\hat{J}_Q^0}(\xi)
{\hat{G}_D}(\xi,\xi'),
\quad{\hat{O}_Q^0}{|_{B}}=0
\]

\noindent and introduced the current
${\hat{J}_Q^0}=Q\hat{R}$. As a result we get the free field integrand

\[
{S_F}[\bar{\phi},\hat{g}]={1\over{2}}\int{d^2}\xi\sqrt{\hat{g}}
\bar{\phi}\hat{\Delta}\bar{\phi}
\]

\noindent plus the non-local functional

\begin{equation}
{\mathcal{F}_D^0}[\hat{g}]={{Q^2}\over{16\pi}}\int{d^2}\xi{d^2}
\xi'\sqrt{\hat{g}(\xi)}\hat{R}(\xi){\hat{G}_D}(\xi,\xi')
\sqrt{\hat{g}(\xi')}
\hat{R}(\xi')\label{11}.
\end{equation}

\noindent Because there is no zero mode ${\hat{G}_D}(\xi,\xi')$ is
Weyl invariant for distinct values of its arguments (coincident values
require regularisation which introduces dependence on the scale of
the metric). Thus, the Weyl anomaly associated with eq. (\ref{11})
is determined by the scaling of the current

\begin{equation}
{\delta_{\rho}}\sqrt{\hat{g}}\hat{R}=\sqrt{\hat{g}}
\hat{\Delta}\rho\label{17}.
\end{equation}
Integrating by parts we find:

\begin{equation}
{\delta_{\rho}}{\mathcal{F}_D^0}={{Q^2}\over{8\pi}}\int{d^2}\xi
\sqrt{\hat{g}}\hat{R}\rho+{{Q^2}\over{8\pi}}
\oint{d}\hat{s}(\xi)\int{d^2}\xi'\rho(\xi){\partial_{\hat{n}}}
{\hat{G}_D}(\xi,\xi')
\sqrt{\hat{g}(\xi')}\hat{R}(\xi')\label{35}.
\end{equation}

\noindent The product of functional determinants resulting from the
integration
over the matter field, the reparametrisations and $\bar \phi$
also varies under a Weyl transformation:

\begin{eqnarray}
&{\delta_{\rho}}\ln\left[{{\left(Det'\hat{\Delta}\right)}^{-(d+1)/2}}
{{\sqrt{Det'{\hat{P}^{\dagger}}\hat{P}}}\over{Vol(CKV)}}\right]=
{{d-25}\over{48\pi}}\int{d^2}\xi\sqrt{\hat{g}}\hat{R}\rho+
{{d-25}\over{24\pi}}\oint{d}\hat{s}{k_{\hat{g}}}\rho+\nonumber\\
&+{{C'}_1}\oint{d}\hat{s}{\partial_{\hat{n}}}\rho+
{{{C'}_2}\over{\varepsilon}}\int{d^2}\xi\sqrt{\hat{g}}\rho+
{{{C'}_3}\over{\sqrt{\varepsilon}}}\oint{d}\hat{s}\rho+
O(\sqrt{\varepsilon})\label{36},
\end{eqnarray}

\noindent where the ${C'}_i$ are dimensionless constants which as before
can be determined exactly.

Ignoring the counterterms for the moment we cancel the bulk local piece
of the Weyl anomaly between eqs. (\ref{35}) and (\ref{36}) if we set

\[
Q=\pm\sqrt{{25-d}\over{6}}.
\]

\noindent Since $\rho$ is an arbitrary infinitesimal Weyl scaling in
the bulk and on the boundary of the surface we also need to deal with the
non-local term and with the local boundary contribution in the
geodesic curvature found respectively in eqs. (\ref{35}) and (\ref{36}).
To do so we have to consider the integration over
the boundary values of the Liouville field.

First we integrate $Y^{\mu}$ and $\eta$. The boundary measures for these
fields are induced by the natural reparametrisation invariant inner
products
on variations of the boundary values:

\[
\left\|\delta{Y}\right\|^2_{\tilde{g}}=\oint{d}\tilde{s}\delta{Y}
\cdot\delta{Y},\quad
\left\|\delta\eta\right\|^2_{\tilde{g}}=\oint{d}\tilde{s}
{{(\delta\eta)}^2}.
\]

\noindent As the formalism is explicitly reparametrisation invariant the
integration over $\eta$ is trivial leading to an overall factor. For the
boundary matter field we find:

\begin{equation}
\int{\mathcal{D}_{\tilde{g}}}Y\exp\left\{-{S_c}[{X_c},\tilde{g}]
\right\}={{\left({{Det'{\tilde{K}_D}}\over{\oint{d}\tilde{s}}}
\right)}^{-d/2}}\int{\prod_{\mu}}d{Y_0^{\mu}}\label{12}.
\end{equation}
Above we took into account the zero mode of the boundary kernel
$\hat{K}_D$. Its existence can be seen by
considering the
eigenvalue problem

\[
\oint{d}\hat{s}(\xi){\hat{K}_D}(\xi,\xi'){\hat{v}_N}(\xi)=
{\hat{\lambda}_N}
{\hat{v}_N}(\xi').
\]

\noindent These eigenfunctions form a complete and
orthonormal set of functions on the boundary:

\[
{\sum_N}{\hat{v}_N}(\xi){\hat{v}_N}(\xi')={\hat{\delta}_B}(\xi-\xi'),
\quad
\oint{d}\hat{s}(\xi){\hat{v}_N}(\xi){\hat{v}_M}(\xi)={\delta_{NM}}.
\]

\noindent Here the boundary delta function is defined by
$\oint{d}\hat{s}(\xi){\hat{\delta}_B}(\xi-\xi')f(\xi)=f(\xi')$. Then the
eigenvalues may be expressed as

\begin{eqnarray*}
&{\hat{\lambda}_N}=\oint{d}\hat{s}(\xi)d\hat{s}(\xi'){\hat{v}_N}(\xi)
{\hat{K}_D}
(\xi,\xi'){\hat{v}_N}(\xi')=\nonumber\\
&=-{1\over{8\pi}}\oint{d}
\hat{s}(\xi)d\hat{s}(\xi'){\hat{v}_N}
(\xi)
{\partial_{\hat{n}}}{\partial_{\hat{n}'}}{\hat{G}_D}(\xi,\xi')
{\hat{v}_N}(\xi').
\end{eqnarray*}
Now define $\hat V_N$ to be the solution of Laplace's equation with
boundary value
$v_N$:

\[
\hat{\Delta}{\hat{V}_N}=0,\quad{\hat{V}_N}{|_B}={\hat{v}_N}.
\]

\noindent This has the solution

\[
{\hat{V}_N}(\xi')=-\oint{d}\hat{s}(\xi){\partial_{\hat{n}}}{\hat{G}_D}
(\xi,\xi'){\hat{v}_N}(\xi),
\]
enabling us to write the eigenvalues as
\[
{\hat{\lambda}_N}={1\over{8\pi}}\oint{d}\hat{s}(\xi){\hat{V}_N}(\xi)
{\partial_{\hat{n}}}{\hat{V}_N}(\xi)=
{1\over{8\pi}}\int{d^2}\sqrt{\hat{g}}{\hat{g}^{ab}}{\partial_a}
{\hat{V}_N}{\partial_b}{\hat{V}_N}.
\]

\noindent Thus ${\hat{\lambda}_N}\geq{0}$ and it is only zero when
$\hat V_N$ is constant. Denoting this solution by $N=0$ and using
the normalisation condition we conclude that $\hat K_D$ has the
zero mode ${\hat{v}_0}={{\left(\oint{d}\hat{s}\right)}^{-1/2}}$.

The determinant in eq. (\ref{12}) will generate a new boundary term
for the Liouville action. This is the gluing anomaly found in \cite{Man}.
The kernel $\tilde{K}_D$ has a boundary
heat kernel which can only be sensitive to
short distance effects, and since the boundary has no intrinsic geometry
it can only be sensitive to
the invariant length of the boundary. As a consequence covariance and
dimensional analysis lead to a contribution to
the Weyl anomaly which can be absorbed into the
cosmological constant counterterm in the invariant world-sheet length of
the boundary.

To cancel the remaining terms in the Weyl anomaly we have to integrate
$\psi$. Just as in the case of $\bar{\varphi}$ we have a non-linear
inner product on variations of $\psi$:

\[
\left\|\delta\psi\right\|^2_{\tilde{g}}=\oint{d}\hat{s}{e^{\psi/2}}
{{(\delta\psi)}^2}.
\]

\noindent We will assume, following David, Distler and Kawai, that
we can use the inner product that is more usual for a quantum
field in the background $\hat{g}_{ab}$,

\[
\left\|\delta\Psi\right\|^2_{\hat{g}}=\oint{d}\hat{s}
{{(\delta\Psi)}^2},
\]

\noindent provided we renormalise ${\psi_0}\to{\alpha_0}{\Psi_0}$,
and $\bar{\psi}\to
{\alpha_B}\bar{\Psi}$ as well as their couplings to 2D quantum
gravity. Note that this means we need to introduce an independent field
renormalisation for $\bar{\varphi}_c$, the component of $\varphi_c$
orthogonal to the zero mode $\psi_0$. According to eq. (\ref{13}), its
explicit
expression in
terms of $\bar{\psi}$ involves a coupling to 2D gravity. Thus we must
also consider ${\bar{\varphi}_c}
\to{\bar{\alpha}_B}
{\bar{\phi}_c}$. This is to be done in each order of the perturbative
expansion in the
length cosmological constant. Note that we have allowed for a different
renormalisation of $\psi_0$ and $\bar{\psi}$. This is because we take
independent bulk and boundary renormalisations and $\psi_0$ is
related to the
zero mode of the Laplacian on closed surfaces that would be generated if
we glued together two disc shaped topologies to obtain a sphere,
corresponding to the inner product of the closed string vacuum with itself.
Thus $\psi_0$ is really associated with the Liouville field in the bulk
and should be renormalised accordingly.

Now when we decompose $\psi$ into $\psi_0$ and $\bar{\psi}$
eq. (\ref{14}) can be rewritten as:

\begin{eqnarray*}
&{S_c^0}[\bar{\psi},{\psi_0},\hat{g}]=-{{d-26}\over{12}}\oint{d}\hat{s}
(\xi)d\hat{s}(\xi')\bar{\psi}(\xi){\hat{K}_{D}}(\xi,\xi')\bar{\psi}
(\xi')-
{{d-26}\over{24\pi}}\oint{d}\hat{s}{k_{\hat{g}}}\bar{\psi}+\nonumber\\
&+{\lambda_1}
\oint{d}\hat{s}{e^{\bar{\psi}+{\psi_0}}}+
{{d-26}\over{48\pi}}\int{d^2}\xi\sqrt{\hat{g}(\xi)}\hat{R}(\xi)
\oint{d}\hat{s}(\xi')
{\partial_{\hat{n}'}}{\hat{G}_D}(\xi,\xi')\bar{\psi}(\xi')-\nonumber\\
&-{{d-26}\over{12}}{\chi_o}{\psi_0}.
\end{eqnarray*}
Introducing the coupling renormalisation parameters $Q_0$, $Q_B$ and
$\bar{Q}_B$ we write the renormalised lowest order boundary action

\begin{eqnarray}
&{S_c^{00}}[\bar{\Psi},{\Psi_0},\hat{g}]={1\over{2}}\oint{d}\hat{s}
(\xi)
d\hat{s}(\xi')\bar{\Psi}(\xi){\hat{K}_D}(\xi,\xi')\bar{\Psi}(\xi')+
\oint{d}\hat{s}
{\hat{H}_D^{00}}{\bar{\Psi}}+\nonumber\\
&+{{{Q_0}{\chi_o}}\over{2}}{\Psi_0},\label{insertp}
\end{eqnarray}

\noindent where we have the current

\begin{equation}
{\hat{H}_D^{00}}(\xi)=-{{Q_B}\over{8\pi}}\int{d^2}\xi'\sqrt{\hat{g}
(\xi')}\hat{R}(\xi')
{\partial_{\hat{n}}}{\hat{G}_D}(\xi,\xi')+{{\bar{Q}_B}\over{8\pi}}
{k_{\hat{g}}}(\xi)\label{16}.
\end{equation}

To integrate this we shift out the linear piece in $\bar{\Psi}$. We
introduce the Green's function of $\hat{K}_D$ defined by

\begin{equation}
\oint{d}\hat{s}(\xi''){\hat{K}_D}(\xi,\xi''){\hat{G}_K}(\xi'',\xi')=
{\hat{\delta}_B}(\xi-\xi')
-{1\over{\oint{d}\hat{s}(\xi''')}}\label{15}.
\end{equation}
The last
term on the right-hand side of eq. (\ref{15}) is necessary to ensure
consistency when the equation is integrated
with respect to $\hat s (\xi)$, since

\[
\oint{d}\hat{s}(\xi){\hat{K}_D}(\xi,\xi')
=0.
\]
Its value is fixed by the zero mode of $\hat{K}_D$ we
have calculated before. Also ${\hat{G}_K}(\xi,\xi')$ is
symmetric in its
arguments and is orthogonal to the constant zero mode

\begin{equation}
\oint{d}\hat{s}(\xi){\hat{G}_K}(\xi,\xi')=0\label{55}.
\end{equation}
Then we can consider the shift
$\bar{\Psi}\to\bar{\Psi}+{\hat{\mathcal{F}}_K^{00}}$ where

\[
{\hat{\mathcal{F}}_K^{00}}(\xi')=\oint{d}\hat{s}(\xi)
{\hat{H}_D^{00}}(\xi)
{\hat{G}_K}(\xi,\xi')
\]

\noindent is also orthogonal to the zero mode. Thus the integration
leads to

\[
\int{\mathcal{D}_{\hat{g}}}(\bar{\Psi},{\Psi_0})\exp\{-
{S_c^{00}}[\bar{\Psi},{\Psi_0},\hat{g}]\}={e^{\mathcal{F}_B^{00}}}
{{\left({{Det'{\tilde{K}_D}}\over{\oint{d}\tilde{s}}}\right)}^{-d/2}}
\int{d}{\Psi_0}{e^{-{Q_0}{\chi_o}{\Psi_0}/2}}
\]

\noindent where

\begin{equation}
{\mathcal{F}_B^{00}}={1\over{2}}\oint{d}\hat{s}(\xi)d\hat{s}(\xi')
{\hat{H}_D^{00}}(\xi){\hat{G}_k}(\xi,\xi'){\hat{H}_D^{00}}(\xi')
\label{31}.
\end{equation}

\noindent The determinant only changes the background renormalisation
counterterm in the world-sheet length. The important contribution to the
Weyl anomaly comes from eq. (\ref{31}). To calculate it we first
need the Weyl transformation associated with eq. (\ref{16}). Using
eq. (\ref{17}) and
the corresponding transformation of the geodesic curvature

\[
{\delta_{\rho}}d\hat{s}{k_{\hat{g}}}={1\over{2}}d\hat{s}
{\partial_{\hat{n}}}\rho,
\]
we take a total derivative and introduce the boundary kernel
$\hat{K}_D$ to find:

\begin{eqnarray}
&{\delta_{\rho}}[d\hat{s}(\xi'){\hat{H}_D^{00}}(\xi')]=
{Q_B}\oint{d}\hat{s}
(\xi)\rho(\xi)d\hat{s}(\xi'){\hat{K}_D}(\xi,\xi')+\nonumber\\
&+{1\over{16\pi}}(
{\bar{Q}_B}-2{Q_B})d\hat{s}(\xi'){\partial_{\hat{n}'}}\rho(\xi')
\label{22}.
\end{eqnarray}
Then eqs. (\ref{15}) and (\ref{16}) lead us to

\begin{equation}
{\delta_{\rho}}{\mathcal{F}_B^{00}}=-{{Q_B^2}\over{8\pi}}
\oint{d}\hat{s}(\xi)\int{d^2}\xi'\rho(\xi){\partial_{\hat{n}}}
{\hat{G}_D}(\xi,\xi')
\sqrt{\hat{g}(\xi)}\hat{R}(\xi')+{{Q_B^2}\over{4\pi}}\oint{d}\hat{s}
{k_{\hat{g}}}\rho\label{32}.
\end{equation}

\noindent Above we have taken ${\bar{Q}_B}=2{Q_B}$ which is a
condition needed to eliminate the contribution associated with the
outward normal derivative of $\rho$:

\[
{{\bar{Q}_B}-2{Q_B}\over 16\pi}\oint{d}\hat{s}(\xi)d\hat{s}(\xi')
{\partial_{\hat{n}}}\rho(\xi){\hat{G}_K}(\xi,\xi'){\hat{H}_D^{00}}
(\xi').
\]

\noindent Also we note that the zero
mode integration defines a net charge selection rule for the
gravitational sector just like in the closed string. This allow us to
ignore the non-local contributions to eq. (\ref{32}) coming from the
zero mode of the kernel $\hat{K}_D$. They will be generated by
eq. (\ref{15}) and by the non-local Weyl anomaly associated with
${\hat{G}_K}(\xi,\xi')$. We find (see Appendix A):

\begin{equation}
{\delta_{\rho}}{\hat{G}_K}(\xi,\xi')=-{1\over{2\oint{d}\hat{s}
(\xi''')}}
\oint{d}\hat{s}(\xi'')
\rho(\xi'')\left[{\hat{G}_K}(\xi'',\xi)+{\hat{G}_K}(\xi'',\xi')
\right]\label{56}.
\end{equation}
This will also contribute when the two points approach
each other. In this case we must also include the contribution coming
from the regularisation of $\hat{G}_K$ at coincident points. We use the
reparametrisation invariant heat kernel

\[
{\hat{G}_{K\varepsilon}}(\xi,\xi')={\int_{\varepsilon}^{\infty}}dt
\left[{\hat{\mathcal{G}}_K}(t,\xi,\xi')-{1\over{\oint{d}\hat{s}
(\xi'')}}
\right]
\]
where $\hat{\mathcal{G}}_K$ satisfies the generalised heat equation

\[
{\partial\over\partial t}{\hat{\mathcal{G}}_K}(t,\xi,\xi')
=\oint{d}\hat{s}(\xi''){\hat{K}_D}(\xi,\xi'')
{\hat{\mathcal{G}}_K}(t,\xi'',\xi'), \quad{\hat{\mathcal{G}}_K}
(0,\xi,\xi')={\hat{\delta}_B}(\xi-\xi').
\]
For coincident arguments the regularisation of the Green's
function is controlled by the small-$t$ behaviour of the
heat kernel which is computable in a standard perturbation
series \cite{Man}. This thus leads to:

\begin{equation}
{\delta_{\rho}}{\hat{G}_{K\varepsilon}}(\xi',\xi')=4\rho(\xi')-
{1\over{\oint{d}\hat{s}(\xi''')}}
\oint{d}\hat{s}(\xi'')
\rho(\xi''){\hat{G}_K}(\xi'',\xi')\label{25}.
\end{equation}

All these non-local contributions always decouple one of the
variables, so they will generate terms in eq. (\ref{32}) which will all
be proportional to the net charge on the whole surface.

To eliminate remaining terms between eqs. (\ref{35}), (\ref{36}) and
(\ref{32}) we need
${Q_B}=Q$. If we finally tune the background cosmological counterterm
contributions to zero we get a Weyl invariant lowest order partition
function. This shows that we need to include the counterterm in the
geodesic curvature because otherwise the finite contribution
coming from the reparametrisation ghosts cannot be eliminated. Of
course in this particular lowest order case we have a null contribution
to the partition function because the net charge,
$\oint{d}\hat{s}(\xi){\hat{H}_D^{00}}(\xi)$, is the topological
background gravity charge which for the disc is
never zero due to the Gauss-Bonnet theorem.
However all the terms we have discussed will persist in the more
complicated expressions that satisfy the charge selection rule.

This analysis still leaves the parameter $Q_0$ undetermined. To
find it we
make the connection with the closed string partition function.
As explained earlier this is obtained by identifying the
arguments of two copies of $Z[Y,\psi,\eta]$ and integrating
over these boundary values.
This corresponds to gluing together two discs along their boundaries
to produce a sphere. The closed string
partition function is

\[
{Z_{closed}}=\int{\mathcal{D}_{\tilde{g}}}(Y,\psi,\eta)
{Z_{open}^1}[Y,\psi,\eta]{Z_{open}^2}[Y,\psi,\eta].
\]

\noindent When we integrate the string field $X^{\mu}$ and the
reparametrisation ghosts in each open string wave functional we find:

\begin{eqnarray*}
&{Z_{closed}}=\int{\mathcal{D}_{\tilde{g}}}(\psi,{\varphi_1},
{\varphi_2})
\exp\{-{S_L}[{\varphi_1},{\hat{g}_1}]-{S_L}[{\varphi_2},
{\hat{g}_2}]\}
\nonumber\\
&{{\left(Det'{\hat{\Delta}_1}\right)}^{-d/2}}
{{\sqrt{Det'{\hat{P}^{\dagger}_1}
{\hat{P}_1}}}\over{{Vol_1}(CKV)}}
{{\left(Det'{\hat{\Delta}_2}\right)}^{-d/2}}
{{\sqrt{Det'{\hat{P}^{\dagger}_2}
{\hat{P}_2}}}\over{{Vol_2}(CKV)}}.
\end{eqnarray*}

\noindent Here the boundary fields $Y^{\mu}$, $\eta$ have already been
integrated and absorbed in the length renormalisation counterterm. The
next step is to perturb in each area renormalisation
counterterm and in the common length cosmological constant. Just like
before we split each field $\varphi_i$, $i=1,2$ in
two independent fields $\varphi_{ci}$, $\bar{\varphi}_i$. In the present
case we only need to consider the lowest order in the perturbative
expansion. Then we have the following decomposition

\begin{eqnarray*}
&{Z_{closed}}={Z_B^{00}}{\bar{Z}_{open}^1}{\bar{Z}_{open}^2}
\nonumber\\
&{{\left(Det'{\hat{\Delta}_1}\right)}^{-d/2}}
{{\sqrt{Det'{\hat{P}^{\dagger}_1}
{\hat{P}_1}}}\over{{Vol_1}(CKV)}}
{{\left(Det'{\hat{\Delta}_2}\right)}^{-d/2}}
{{\sqrt{Det'{\hat{P}^{\dagger}_2}
{\hat{P}_2}}}\over{{Vol_2}(CKV)}}
\end{eqnarray*}

\noindent where the boundary partition function is

\[
{Z_B^{00}}=\int{\mathcal{D}_{\tilde{g}}}(\psi,{\psi_0})\exp\{-
{S_B^{00}}[\psi,{\psi_0},\hat{g}]\}.
\]

\noindent Above we have used the simple property that the outward normal
derivative of one of the open surfaces is just the inward normal
derivative of the other at the common boundary, plus the Gauss-Bonnet
theorems given in eqs. (\ref{5}) and (\ref{18}) to find the
boundary action:

\[
{S_B^{00}}(\bar{\psi},{\psi_0},\hat{g})=-{{d-26}\over{12}}
\oint{d}\hat{s}d
{\hat{s}'}\bar{\psi}(\xi){\hat{K}_{D}}(\xi,\xi')\bar{\psi}(\xi')-
{{d-26}\over{12}}{\chi_c}{\psi_0}.
\]

Next we renormalise the fields and
their couplings to 2D gravity to consider canonical measures in the
background
$\hat{g}_{ab}$. When we integrate $\bar{\phi}_i$ we get the same Weyl
anomaly
for each field and again using the property of the normal
derivative at the common boundary we can easily see
that the boundary contributions
cancel up to the usual length renormalisation counterterm, leading to
${Q_i}=Q$, $i=1,2$, where the $Q_i$ define the renormalisation of the
coupling of the $\varphi_i$ to the scalar curvature $\hat{R}_i$. The
boundary integration is just equal to the zero mode
charge selection rule. If ${Q_0}=Q$ that is exactly the selection rule
we get for the closed string.

\subsubsection{Anomaly cancellation for Liouville field renormalisation}

So far we have only been able to determine parameters associated with
the renormalisation of the couplings to 2D quantum gravity. To go
further and calculate the Liouville field renormalisation we need to
consider higher orders in the Coulomb gas perturbative expansion. In the
case of the couplings we have seen that the
renormalised central charge of the conformally extended Liouville field
theory is exactly the same as the corresponding central charge of the
same theory on a closed surface. We have also proved that the boundary
couplings are fixed by this value of the central charge. We have seen
that this is all a consequence of the quantum Weyl invariance of the
theory. Now we want to find out if the bulk field renormalisation is
equal to the corresponding closed string parameter and if the boundary
field renormalisation is actually the same as its bulk counterpart as it
should happen when we interpret the Liouville field as
an arbitrary Weyl scaling defined everywhere on the surface including
its boundary. As Symanzik's work makes clear,
this is not something we should take
for granted. We will now show that this is also a consequence
of the quantum Weyl invariance assumed for the theory.

We start with the case where we have a single Liouville vertex operator
on the bulk:

\begin{equation}
\int{d^2}\xi\sqrt{\hat{g}}{e^{{\alpha_0}{\Psi_0}+\alpha\bar{\phi}+
{\bar{\alpha}_B}
{\bar{\phi}_c}}}\label{24}.
\end{equation}

\noindent In this case we find the following action for $\bar{\phi}$

\[
{S_L^{1}}\left[\bar{\phi},\hat{g}\right]={1\over{8\pi}}\int{d^2}\xi
\sqrt{\hat{g}}\left({1\over{2}}
\bar{\phi}\hat{\Delta}\bar{\phi}+{\hat{J}_Q^1}\bar{\phi}\right)
\]

\noindent where we need the current

\[
{\hat{J}_Q^1}(\xi)=Q{\hat{R}}(\xi)-8\pi\alpha{{{\delta^2}(\xi-\xi')}
\over{\sqrt{\hat{g}(\xi)}}}.
\]

\noindent By shifting $\bar{\phi}$ we generate the functional:

\begin{equation}
{\mathcal{F}_D^1}[\hat{g}]={\mathcal{F}_D^0}[\hat{g}]-\alpha{Q}
\int{d^2}
\xi\sqrt{\hat{g}(\xi)}{\hat{R}}(\xi){\hat{G}_D}(\xi,\xi')+4\pi
{\alpha^2}
{\hat{G}_D}(\xi',\xi')\label{20}.
\end{equation}

\noindent On the other hand we also find the following renormalised
boundary action

\begin{eqnarray*}
&{S_c^{10}}[\bar{\Psi},{\Psi_0},\hat{g}]={1\over{2}}\oint{d}\hat{s}
(\xi)
d\hat{s}(\xi')\bar{\Psi}(\xi){\hat{K}_D}(\xi,\xi')\bar{\Psi}(\xi')+
\oint{d}\hat{s}
{\hat{H}_D^{10}}{\bar{\Psi}}+\nonumber\\
&+\left({{{Q_0}{\chi_o}}\over{2}}-{\alpha_0}\right)
{\Psi_0}
\end{eqnarray*}

\noindent where we have introduced the current

\[
{\hat{H}_D^{10}}(\xi)={\hat{H}_D^{00}}(\xi)+
{\bar{\alpha}_B}{\partial_{\hat{n}}}{\hat{G}_D}(\xi,\xi').
\]

\noindent In this case we get

\begin{eqnarray}
&{\mathcal{F}_B^{10}}={\mathcal{F}_B^{00}}+{\bar{\alpha}_B}\oint{d}
\hat{s}
(\xi)d\hat{s}(\xi'')
{\hat{H}_D^{00}}(\xi){\hat{G}_K}(\xi,\xi''){\partial_{\hat{n}''}}
{\hat{G}_D}(\xi'',\xi')
+\nonumber\\
&+{1\over{2}}{\bar{\alpha}_B^2}\oint{d}\hat{s}(\xi)
d\hat{s}(\xi''){\partial_{\hat{n}}}{\hat{G}_D}(\xi,\xi')
{\hat{G}_K}(\xi,\xi'')
{\partial_{\hat{n}''}}{\hat{G}_D}(\xi'',\xi')\label{23}.
\end{eqnarray}

To analyse the anomaly cancellation in this order of the perturbative
expansion we first recall that although ${\hat{G}_D}(\xi,\xi')$ is
Weyl invariant for distinct values
of its arguments, at coincident points it
requires regularisation which introduces dependence on the scale of the
metric. To calculate the correspondent Weyl transformation we represent
${\hat{G}_D}(\xi,\xi')$ in terms of the
Green's function $\hat{G}(\xi,\xi')$ considered on the whole plane

\[
{\hat{G}_D}(\xi,\xi')=\hat{G}(\xi,\xi')-{\hat{H}_D}(\xi,\xi'),
\]
where ${\hat{H}_D}(\xi,\xi')$ satisfies the boundary-value problem:

\[
\hat{\Delta}{\hat{H}_D}(\xi,\xi')=0,\quad{\hat{H}_D}
(\xi,\xi'){|_{\xi'\in B}}
=\hat{G}(\xi,\xi'){|_{\xi'\in B}}.
\]
When $\xi=\xi'$ is on the bulk ${\hat{H}_D}(\xi,\xi)$ is Weyl
invariant. Also on the whole plane there is no zero mode. Thus the Weyl
transformation of
${\hat{G}_D}(\xi,\xi)$ is just given by the corresponding well known
local change of $\hat{G}(\xi,\xi)$ \cite{Pol,Pho}:

\begin{equation}
{\delta_{\rho}}{\hat{G}_{D\varepsilon}}(\xi,\xi)={{\rho(\xi)}\over
{4\pi}},\quad\xi
\not\in{B}\label{21} .
\end{equation}
Then applying eqs. (\ref{17}), (\ref{21}) we conclude that the Weyl
anomaly of eq. (\ref{20}) is given by

\[
{\delta_{\rho}}{\mathcal{F}_D^1}={\delta_{\rho}}{\mathcal{F}_D^0}-
\alpha{Q}\oint{d}\hat{s}(\xi)\rho(\xi){\partial_{\hat{n}}}{\hat{G}_D}
(\xi,\xi')+({\alpha^2}-\alpha{Q})\rho(\xi').
\]

On the other hand, ignoring the non-local zero mode contributions
which are all proportional to the net charge on the whole surface
given in this order by $\oint{d}\hat{s}{\hat{H}_D^{10}}$, we use
eq. (\ref{22}) and ${\bar{Q}_B}=2{Q_B}$ to find the Weyl anomaly of
eq. (\ref{23}):

\[
{\delta_{\rho}}{\mathcal{F}_D^{10}}={\delta_{\rho}}
{\mathcal{F}_D^{00}}+
{\bar{\alpha}_B}{Q_B}\oint{d}\hat{s}(\xi)\rho(\xi)
{\partial_{\hat{n}}}
{\hat{G}_D}(\xi,\xi').
\]

Thus we can easily see that to ensure Weyl invariance at the quantum
level we must further set ${Q_B}=Q$,
${\bar{\alpha}_B}=\alpha$ and

\[
1-\alpha{Q}+{\alpha^2}=0.
\]

\noindent Here we took into account the contribution to the Weyl
anomaly of the $\sqrt{\hat{g}}$ present in eq. (\ref{24}).
Introducing the value of $Q$ we find:

\[
{\alpha_{\pm}}={1\over{2\sqrt{6}}}\left(\sqrt{25-d}\pm\sqrt{1-d}
\right).
\]

As we noted previously, these renormalised parameters only cancel
the local contributions to the Weyl anomaly. As in the lowest
order case we have to assume the charge selection rule
associated with the zero mode integration to eliminate the
non-local pieces. To find the renormalised
parameters of the charge selection rule we need to glue the
two discs to form a sphere enabling us to use the closed string
result. We
already know the value of $Q_0$ but now we also want the value
of $\alpha_0$. The calculation goes exactly as before, all boundary
contributions cancel out up to the length renormalisation counterterm
and we find a zero mode integral which corresponds to a
closed string selection rule with two bulk vertex operator charges
$\alpha_0$ and a background gravity charge $Q_0=Q$.
This implies that ${\alpha_0}=\alpha$ as expected.

With this calculation we are able to guarantee Weyl invariance at the
quantum level for insertions of arbitrary numbers of gravitational
Liouville vertex operators in the bulk. To see what happens when
operators are
inserted on the boundary let us consider the simplest case of just one
such
operator,

\begin{equation}
\oint{d}\hat{s}{e^{{\alpha_0}{\Psi_0}/2+{\alpha_B}\bar{\Psi}/2}}
\label{26}.
\end{equation}

In this case only the boundary
integration over $\psi$ gets changed. The renormalised boundary
Liouville action is

\begin{eqnarray*}
&{S_c^{01}}[\bar{\Psi},{\bar{\Psi}_0},\hat{g}]={1\over{2}}\oint{d}
\hat{s}(\xi)d\hat{s}(\xi')\bar{\Psi}(\xi){\hat{K}_D}(\xi,\xi')
\bar{\Psi}(\xi')+
\oint{d}\hat{s}{\hat{H}_D^{01}}\bar{\Psi}+\nonumber\\
&+\left({{{Q_0}{\chi_o}}
\over{2}}-
{{\alpha_0}\over{2}}\right){\bar{\Psi}_0}
\end{eqnarray*}

\noindent where we have introduced the current

\[
{\hat{H}_D^{01}}(\xi)={\hat{H}_D^{00}}(\xi)-{{\alpha_B}\over{2}}
{\hat{\delta}_B}(\xi-\xi').
\]
The relevant functional is now:

\[
{\mathcal{F}_B^{01}}={\mathcal{F}_B^{00}}-{{\alpha_B}\over{2}}\oint{d}
\hat{s}(\xi){\hat{H}_D^{00}}(\xi){\hat{G}_K}(\xi,\xi')+
{{\alpha_B^2}\over{8}}{\hat{G}_K}(\xi',\xi').
\]
To find out our last renormalised parameter $\alpha_B$ we
need the local Weyl transformation of $\hat{G}_K$ at coincident points
given in
eq. (\ref{25}). Thus the local anomaly vanish if all the other
parameters keep their previous values and
$1/2-{\alpha_B}Q/2+{\alpha_B^2}/2=0$,
where the $1/2$ term comes from the Weyl transformation of $d\hat{s}$
in eq. (\ref{26}). Thus ${\alpha_B}=\alpha$.

Since as before the non-local contributions cancel due to the charge
selection rule this result shows that the full perturbative expansion
is Weyl
invariant at the quantum level for the values of the renormalised
parameters found. Whenever we couple distinct Liouville vertex
operators in higher orders
there are no additional Weyl anomalous contributions.

\subsubsection{Comments}

Our results for the non-critical open
string show that the gravitational sector can be interpreted as a
conformally extended boundary Liouville field theory. In this picture
$Q$ defines the central charge of the
Liouville theory ${c_{\phi}}=1+6{Q^2}$, which has its value fixed by
demanding that it cancels the central charges of the matter and ghost
systems ${c_M}+{c_{gh}}=d-26$. Thus the central charge of the theory
with boundary is equal to the central charge of the theory without
boundary. This is to be expected since anomalies are local effects.

We have interpreted the Liouville field as an arbitrary Weyl scaling
all over the open surfaces. Then we found that the value of $\alpha$ is
exactly right to define a Liouville vertex operator
$\int{d^2}\xi\sqrt{\hat{g}}{e^{\alpha\phi}}$ of zero conformal weight.
On the extended field theory it corresponds to a primary field
$:{e^{\alpha\phi}}:$ of weight $(1,1)$. As expected $\alpha$ has the
same value it takes when the surfaces are closed. We also found the
right value for
$\alpha_B$ in the sense that the boundary vertex operator
$\oint{d}\hat{s}{e^{{\alpha_B}\phi/2}}$ has zero conformal weight
corresponding to the boundary primary field $:{e^{{\alpha_B}\phi/2}}:$
of conformal weight $(1/2,1/2)$. This means that the renormalisation
of the Liouville field is the same all over the surface and is equal
to the renormalisation on the closed surface as it should be.

We are now in a position to see that Dirichlet boundary conditions on
the Liouville field imply that the metric is discontinuous as the
boundary is approached. In this case the calculation
stops at $S^{00}_c$ in eq. (\ref{insertp}), since we do not integrate over
boundary values of the Liouville field, but leave them fixed. The Weyl
anomaly of
eq. (\ref{36}) must now be cancelled by the Weyl transform of $S^{00}_c$,
together with a shift in the boundary value of the Liouville field,
$\Psi$. This fixes
the latter to be $\delta\Psi=-Q\rho$. Now the full metric is a
reparametrisation
of ${\hat{g}}_{ab}{e^{\alpha\Psi}}$, which should be invariant under
this simultaneous Weyl transformation on ${\hat{g}}_{ab}$ and shift in
$\Psi$, since the separation into reference metric and Liouville field
is arbitrary. However it is not because $Q\ne 1/\alpha$, as the correct
relation, $1-\alpha Q +\alpha^2=0$, has an extra quantum piece.
One way out of this would be
to assume that the Liouville field is renormalised differently in the
bulk and on the boundary, a phenomenon that occurs in $\phi^4$ theory in
four dimensions, \cite{SYM}. However, this implies that the metric is
discontinuous as the boundary is approached, and also that the
functionals obtained
by imposing Dirichlet boundary conditions cannot be sewn together to
make closed surface functionals.

As for the closed string we also find the
need to restrict the
validity of the approach to target space dimensions $d\leq1$. Only
in this way we have real renormalised parameters such that
$e^{\alpha\phi}$ and $e^{{\alpha_B}\phi/2}$ can be interpreted as real
Weyl scalings for a real scalar renormalised Liouville field
$\phi$. From this we can see that our results extend very
naturally those found for the closed string by David, Distler
and Kawai. Since the analysis is fully local and we can choose the
moduli integration measure to be independent of the conformal factor of
the metric our results also generalise immediately to higher genus
Riemann surfaces with just one boundary. Clearly more general boundary
structures can also be considered. Here for simplicity we have just
analysed the random surfaces one loop functional defined in
euclidean space. Our results
hold for an arbitrary number of loops. We also may consider non-smooth
boundaries \cite{Man}.

\subsubsection{Tachyon gravitational dressings}

Since this formalism is only valid for $d\leq1$ the string
serves as a toy model for the more realistic $c\leq{1}$ minimal
series of boundary conformal field theories \cite{JC}. With this in
mind let us see how a bulk tachyon vertex operator gets dressed
by the gravitational sector. Taking an n-point function of
bulk tachyons with momentum $p_j$ such that the momenta sum to
zero we can easily see by following the same path of
calculations that the operator
$\int{d^2}{\xi_j}\sqrt{\tilde{g}_j}{e^{i{p_j}\cdot{X}({\xi_j})}}$
gets dressed to
$\int{d^2}{\xi_j}\sqrt{\hat{g}_j}{e^{{\gamma_j}\phi}}{e^{i{p_j}
\cdot{X}
({\xi_j})}}$, where quantum Weyl invariance demands

\[
{\Delta_j^0}-{\gamma_j}({\gamma_j}-Q)=1,\quad{\Delta_j^0}=
{p_j^2}.
\]

\noindent The above equation shows that the dressed bulk tachyon vertex
operator has zero conformal weight. The primary Liouville field $:
{e^{{\gamma_j}\phi}}:$ dresses the tachyon field $:{e^{i{p_j}\cdot{X}(
{\xi_j})}}:$ in such a way that $:{e^{{\gamma_j}\phi}}{e^{i{p_j}
\cdot{X}({\xi_j})}}:$ has conformal weight $(1,1)$. Note that if we
solve for
$\gamma_j$ in terms of $Q$ and $p_j^2$ we get ${\gamma_j}=(1/2)
\left[Q\pm\sqrt{{Q^2}+4({p_j^2}-1)}\right]$. Just as $\alpha$
should be real for an arbitrary real Weyl scaling so should be
$\gamma_j$. This
implies that $d\leq1$ and ${p_j^2}\geq{0}$.

If we now consider a boundary tachyon vertex operator $\oint{d}
{\hat{s}_j}{e^{i{p_j}\cdot{X}({\xi_j})/2}}$ we may follow the
same steps to find the dressed operator $\oint{d}{\hat{s}_j}
{e^{{\gamma_j}\phi/2+i{p_j}\cdot{X}({\xi_j})/2}}$ where $\gamma_j$
satisfies the same equation as the its bulk
counterpart. It means that the dressed field
$:{e^{i{p_j}\cdot{X}({\xi_j})/2}}
{e^{{\gamma_j}\phi/2}}:$ considered in the boundary of the surface
has conformal weight (1/2,1/2) as a consequence of the quantum Weyl
invariance of the theory.

\subsection{Neumann boundary conditions}

The choice of boundary conditions always depends on the
specific physical applications we have in mind. So far we have argued
that in a proper coupling to 2D quantum gravity, the boundary conditions
on the Liouville field have to be such that it
can be
interpreted as an arbitrary Weyl scaling on the whole surface and not
just on its interior. As we said this rules out Dirichlet boundary
conditions but we are free to choose Neumann boundary conditions for the
conformal
factor. To see what happens in this case let us for
simplicity take
also Neumann boundary conditions on the matter field
${\partial_{\tilde{n}}}{X^{\mu}}=0$ and on
the reparametrisation ghosts $\tilde{n}\cdot\delta\theta=0$. Consider
first the partition function. We can then follow the same reasoning as
in the case of free boundary conditions with much more ease because the
Neumann boundary condition simply eliminates the most part of the
boundary contributions we had to worry about before. Thus we write the
following renormalised Liouville action

\begin{eqnarray*}
&{S_L^N}\left[\phi,\hat{g}\right]={1\over{8\pi}}\int{d^2}\xi
\sqrt{\hat{g}}\left({1\over{2}}
\phi\hat{\Delta}\phi+Q\hat{R}\phi\right)+{{\bar{Q}_B}\over{8\pi}}
\oint{d}\hat{s}{k_{\hat{g}}}\phi+\nonumber\\
&+{\mu_2^2}\int{d^2}
\sqrt{\hat{g}}{e^{\alpha\phi}}+{\lambda_2}\oint{d}\hat{s}
{e^{{\alpha_B}\phi/2}}.
\end{eqnarray*}

\noindent Here $Q$, $\bar{Q}_B$ refer to coupling renormalisation and
$\alpha$,
$\alpha_B$ are its field renormalisation counterparts. To ensure quantum
Weyl invariance (see Appendix B) we must satisfy the charge conservation
selection rule, tune the local reference
counterterms to zero,
set ${\bar{Q}_B}=2Q$, ${\alpha_B}=\alpha$ plus
$Q=\pm\sqrt{(25-d)/6}$ and $1-\alpha{Q}+{\alpha^2}=0$.

Starting from a general open string bulk tachyon amplitude it is clear
that
we may follow the steps of the partition function calculation to find the
equation for the gravitational dressing of the bulk tachyon vertex
operator. A tachyon vertex operator with momentum $p_j$ gets dressed by
the coupling to 2D quantum gravity
$\int{d^2}{\xi_j}\sqrt{\hat{g}}{e^{{\gamma_j}\phi}}
{e^{i{p_j}\cdot{X}({\xi_j})}}$, where
${\Delta_j^0}-{\gamma_j}({\gamma_j}-Q)=1$, ${\Delta_j^0}=
{p_j^2}$. For the boundary tachyon vertex
operator the
coupling to gravity leads to the dressed operator
$\oint{d}{\hat{s}_j}{e^{{\gamma_j}\phi/2+
i{p_j}\cdot{X}({\xi_j})/2}}$ of zero weight, where $\gamma_j$ satisfies
the same equation as the its bulk counterpart.

Thus so far we conclude that our results are exactly the same for
Neumann and for free boundary conditions on the Liouville field.

\section{Critical exponents and the saddle point limit}

\subsection{The open string susceptibility and Yang-Mills
Feynman mass exponents}

Let us consider again the case of free boundary conditions and start
with Polyakov's sum over random surfaces with the topology of a
disc. Generalising the
closed string case \cite{ZCKT} the quantum
partition function may now be written as an integral
of the partition function for surfaces constrained to have fixed area,
$A$, and perimeter, $L$, $\Gamma (A,L)$:

\[
Z={\int_0^{+\infty}}\Gamma(A,L){e^{-{\mu_0^2}A-{\lambda_0}L}}dAdL.
\]

\noindent After integrating out the matter and reparametrisation ghost
fields
we renormalise the Liouville field and its couplings to 2D gravity to
find the following integral for $\Gamma (A,L)$:

\begin{eqnarray}
&\Gamma(A,L)=\int{\mathcal{D}_{\hat{g}}}(\bar{\Psi},\bar{\phi})
d{\bar{\Psi}_0}{{\left(\oint{d}\hat{s}\right)}^{1/2}}{e^{-{S_c^{00}}
[\bar{\Psi},{\bar{\Psi}_0},\hat{g}]-{\bar{S}^0}
[\bar{\phi},\hat{g}]}}\delta\left(\int{d^2}\xi\sqrt{\hat{g}}
{e^{\alpha\phi}}-A\right)\nonumber\\
&\delta\left(\oint{d}\tilde{s}
{e^{\alpha/2\phi}}-L\right)\label{28}.
\end{eqnarray}

\noindent Here we have factored out the cosmological constant
counterterms left over from the renormalisation process. Note that in
this process the initially infinite constants $\mu_0^2$ and $\lambda_0$
are changed into the finite constants $\mu_2^2$ and $\lambda_2$. As
discussed before we have set ${\nu_2}=0$.

To calculate the critical exponents we apply David, Distler and Kawai's
scaling argument \cite{DDK}. Consider the shift of the integration
variable $\phi$
by a constant
$\phi\to\phi+\rho/\alpha$. Since we keep $\hat{g}_{ab}$
fixed our functional integral should scale. Recall that
the
theory is invariant under
arbitrary scalings of the reference metric once we have integrated
$\phi$. So, it is only invariant under a shift of the integration
variable
provided this is compensated by a Weyl transformation of the reference
metric. Because we consider a translational invariant quantum measure in
eq. (\ref{28}), the scaling behavior is determined by the change in the
action $S\equiv{S_c^{00}}[\bar{\Psi},{\bar{\Psi}_0},\hat{g}]+{\bar{S}^0}
[\bar{\phi},\hat{g}]$, and in the delta functions which are used to fix
the area $A$ and the perimeter $L$ of the surface. Being the shift
constant only the zero mode
$\bar{\Psi}_0$ is actually changed. Thus the shift in the action is

\[
S\to{S}+{{Q{\chi_o}}\over{2\alpha}}\rho,
\]
and the shifts in the delta functions are

\begin{eqnarray*}
&\delta\left(\int{d^2}\xi\sqrt{\hat{g}}{e^{\alpha\phi}}-A\right)\to
{e^{-\rho}}\delta\left(\int{d^2}\xi\sqrt{\hat{g}}{e^{\alpha\phi}}-
{e^{-\rho}}A\right),\\
&\delta\left(\oint{d}\hat{s}{e^{\alpha\phi/2}}-L\right)\to{e^{-\rho/2}}
\delta\left(\oint{d}\hat{s}{e^{\alpha\phi/2}}-{e^{-\rho/2}}L\right).
\end{eqnarray*}
Then we get the following scaling law:

\begin{equation}
\Gamma(A,L)={e^{-{\rho\over{2}}\left({{{\chi_o}Q}\over
{\alpha}}+3\right)}}
\Gamma\left({e^{-\rho}}A,{e^{-\rho/2}}L\right)\label{42}.
\end{equation}

To be able to introduce critical exponents we have to define the
partition function for fixed area $A$, $\Sigma(A)$,
and the partition function for fixed perimeter $L$, $\Omega(L)$.
Factoring out the appropriate counterterms we write:
\[
\Sigma(A,{\lambda_2})=\int\Gamma(A,L){e^{-{\lambda_2}L}}dL,\quad
\Omega(L,{\mu_2^2})=\int\Gamma(A,L){e^{-{\mu_2^2}A}}dA.
\]
Then, from eq. (\ref{42}), we get:

\begin{equation}
\Sigma(A,{\lambda_2})={e^{-\rho\left({{{\chi_o}Q}\over{2\alpha}}+
1\right)}}
\Sigma\left({e^{-\rho}}A,{\lambda_2}{e^{\rho/2}}\right)\label{50},
\end{equation}

\begin{equation}
\Omega(L,{\mu_2^2})={e^{-{\rho\over{2}}\left({{{\chi_o}Q}\over{\alpha}}+
1\right)}}
\Omega\left({e^{-\rho/2}}L,{\mu_2^2}{e^{\rho}}\right)\label{51}.
\end{equation}

The open string susceptibility exponent is defined just like in the
closed string. In the case
${\lambda_2}=0$ we can continue to use the scaling argument. As
$A\to+\infty$:

\[
\Sigma(A)\sim{A^{\sigma({\chi_o})-3}}.
\]
and

\[
\sigma({\chi_o})=2-{{{\chi_o}Q}\over{2\alpha}}.
\]

\noindent The last result is just the expected open string version of the
closed string critical exponent. If we take the positive root for $Q$ and
the corresponding negative one for $\alpha$ we find that in the
semi-classical
limit $d\to-\infty$:

\[
\sigma({\chi_o})={{d-19}\over{12}}{\chi_o}+2.
\]

For the open string we can also consider the asymptotic limit
$L\to+\infty$ and introduce a mass critical exponent in close analogy
with the the asymptotic limit $A\to+\infty$. Here we
take ${\mu_2^2}=0$. This case was considered by Durhuus,
Olesen and Petersen \cite{LSW} in connection with the calculation of
the Wilson loop quark-antiquark potencial. We define
$\omega({\chi_o})$ by

\[
\Omega(L)\sim{L^{\omega({\chi_o})-3}}.
\]

\noindent Thus we find

\[
\omega({\chi_o})=2-{{{\chi_o}Q}\over{\alpha}},
\]

\noindent to which we associate the semi-classical limit

\[
\omega({\chi_o})={{d-19}\over{6}}{\chi_o}+2.
\]

We can interpret of $\omega({\chi_o})$ in the context
of Yang-Mills gluon dynamics. To see this first note that the wave
functional given in eq. (\ref{29}) models the Wilson loop ,W, for
Yang-Mills theory \cite{OAL,LSW}.
Consider the first quantised functional integral representing the
propagator
of a particle of mass $\lambda_2$ moving under the influence of a
Yang-Mills
field. At coincident
points its trace is a gauge invariant expression
\[
\int{\mathcal{D}}Y\,tr\,P\,e^{-{\lambda_2}\oint{d}\tilde{s}-
\oint dY\cdot{\cal A}}
=tr\,G_{\cal A}(x,x).
\]
If this is averaged over the Yang-Mills field we get
\[
<tr\,G_{\cal A}(x,x)>_{\cal A}=\int{\mathcal{D}Y}\,e^{-{\lambda_2}
\oint{d}\tilde{s}}\,W
=\int dL\,e^{-\lambda_2L}\int{\mathcal{D}Y}\,\delta(L-\oint{d}\tilde{s})
\,W
\]
but this last functional integral is just what we mean by $\Omega$.
Substituting
the form that holds for $\mu_2^2=0$ we get
\[
<tr\,G_{\cal A}(x,x)>_{\cal A}
\propto{{\lambda_2}^{{\chi_o}Q/\alpha}}={{\lambda_2}^
{2-\omega({\chi_o})}},
\]
valid for small $\lambda_2$, corresponding to large $L$.
Thus $\omega({\chi_o})$ is the critical exponent associated with the
Feynman propagator of a test particle which interacts with the
Yang-Mills gauge fields.

So far we have expanded the cosmological terms so as to linearise
the contribution of the exponential terms to the action. We will
now discuss a different approach based on the semi-classical expansion.

\subsection{The saddle point expansion}

When we consider the partition function with free boundary conditions
and integrate the matter and ghost fields in the conformal gauge
${\tilde{g}_{ab}}={e^{\varphi}}{\hat{g}_{ab}}$, the result is

\[
\Gamma(A,L)=\int{\mathcal{D}_{\tilde{g}}}(\psi,\varphi){e^{-{S_L}
[\varphi,
\hat{g}]}}\delta\left(\int{d^2}\xi\sqrt{\hat{g}}{e^{\varphi}}-A\right)
\delta\left(\oint{d}\hat{s}{e^{\varphi/2}}-L\right)
\]

\noindent where the Liouville action is given by

\[
{S_L}[\varphi,\hat{g}]={{26-d}\over{48\pi}}\int{d^2}\xi\sqrt{\hat{g}}
\left({1\over{2}}{\hat{g}^{ab}}{\partial_a}\varphi{\partial_b}\varphi+
\hat{R}\varphi\right)+{{26-d}\over{24\pi}}\oint{d}\hat{s}{k_{\hat{g}}}
\varphi.
\]

Representing the delta functions by integrals over (imaginary) Lagrange
multipliers $p,q,$ gives the Euclidean action

\begin{eqnarray*}
&{S_L}[\varphi,\hat{g},p,q]={{26-d}\over{48\pi}}\int{d^2}\xi\sqrt{\hat{g}}
\left({1\over{2}}{\hat{g}^{ab}}{\partial_a}\varphi{\partial_b}\varphi+
\hat{R}\varphi\right)+{{26-d}\over{24\pi}}\oint{d}\hat{s}{k_{\hat{g}}}
\varphi-\nonumber\\
&-p\left(\int{d^2}\xi\sqrt{\hat{g}}{e^{\varphi}}-A\right)-q\left(\oint
{d}\hat{s}{e^{\varphi/2}}-L\right).
\end{eqnarray*}

\noindent This action is invariant under M\"obius transformations on
the upper half-plane, i.e.
$SL(2,\mathcal{R})$ invariant. These
transformations preserve the conformal gauge, mapping the upper half-plane
onto itself $\omega\to\omega'=(a\omega+b)/(c\omega+d)$,
$\varphi(\omega,\bar{\omega})\to\varphi(\omega',\bar{\omega}')+2\ln
|d\omega'/d\omega|$, where $a,b,c,d\in\mathcal{R}$ and $ad-bc=1$. It will
be more convenient to work on the unit disc obtained from the upper
half-plane
by the complex M\"obius transformation
$\omega\to{z}=(i-\omega)/(i+\omega)$,
$\varphi(\omega,\bar{\omega})\to\varphi(z,\bar{z})+2\ln
|dz/d\omega|$. To get the correspondent invariance on the unit disc we
map the
$SL(2,\mathcal{R})$ transformation. The result is the conformal mapping
of the unit disc onto itself
$z\to{z'}=\exp(i{\theta_0})(z+{c_0})/(1+{\bar{c}_0}z)$,
$\varphi(z,\bar{z})\to\varphi(z',\bar{z}')+2\ln
|dz/d\omega|$, where ${\theta_0}\in\mathcal{R}$ and $|{c_0}|<1$.

In the saddle point approximation we expand around the solution of the
following classical problem:

\[
\tilde{R}=\eta,\quad{2}{k_{\tilde{g}}}=k,\quad
\int{d^2}\xi\sqrt{\tilde{g}}=A,\quad\oint{d}\tilde{s}=L
\]

\noindent where $\eta=p\gamma$, $\gamma=48\pi/(26-d)$ and $2k=q\gamma$.
This is a boundary-value problem for the Liouville field of the
conformal gauge ${\tilde{g}_{ab}}={e^{\varphi}}
{\hat{g}_{ab}}$. The classical field $\varphi_c$ must satisfy the
Liouville equation

\[
\hat{R}+\hat{\Delta}\varphi_c=\eta{e^{\varphi_c}},\quad\int{d^2}\xi
\sqrt{\hat{g}}{e^{\varphi_c}}=A
\]

\noindent subject to the boundary condition \cite{GNM}:

\[
2{k_{\hat{g}}}+{\partial_{\hat{n}}}\varphi_c=k{e^{\varphi_c/2}},
\quad\oint{d}\hat{s}{e^{\varphi_c/2}}=L.
\]

\noindent Here $\eta$ and $k$ are not independent.
Applying the Gauss-Bonnet theorem gives $\eta{A}+kL=4\pi$.

Let us now solve
this problem on the unit disc. In the polar coordinates
$z=\rho{e^{i\theta}}$, $\rho\in[0,1]$, $\theta\in[0,2\pi]$ we assume
that $\varphi_c$ only depends on $\rho$ to find the
following solution:

\[
{\varphi_c}(\rho)=2\ln{{2A}\over{L}}{{\left[1+\left({{4\pi{A}}
\over{L^2}}-1
\right){\rho^2}\right]}^{-1}}.
\]

\noindent Due to the $\theta$ independence this is the metric of a
spherical
cap of length $L$ and area $A$. Note that ${\rho^2}{e^{\varphi_c}}>0$
leads to ${L^2}<4\pi{A}$. We also note that
${\eta_c}=8\pi/A[1-{L^2}/(4\pi{A})]$, ${k_c}=2L/A(1-2\pi{A}/{L^2})$.

The saddle point tree level approximation is given by the classical
functional ${e^{-{S_L}[{\varphi_c},\hat{g},{p_c},{q_c}]}}$. Introducing
the new coordinate $\varrho$ such that $\rho$ is given by
$\rho=\tan(\varrho/2)/
\sqrt{4\pi{A}/{L^2}-1}$, we obtain

\[
\Gamma(A,L)={{\left({A\over{L}}\right)}^{(d-26)/6}}
{{\left({{L^2}\over{4\pi{A}}}\right)}^{2{L^2}/A\gamma}}
{e^{-2{L^2}/A\gamma}}.
\]

\noindent In the semi-classical limit $d\to-\infty$ we get

\[
\Gamma(A,L)={e^{d/12\rho}}\Gamma\left({e^{-\rho}}A,{e^{-\rho/2}}L
\right).
\]

\noindent If we take the branch $\alpha_{-}$, ${\chi_o}=1$ and the
limit $d\to-\infty$ we reproduce this scaling law from eq. (\ref{42}) so
that in the case of the disc topology both methods match in the
asymptotic limit $A\to+\infty$,
$L\to+\infty$ such that $A/{L^2}\to{const}$.

If we go to one loop we must consider

\[
\Gamma(A,L)={e^{-{S_L}[{\varphi_c},\hat{g},{p_c},{q_c}]}}\int
{\mathcal{D}_{g_c}}(\phi,\chi)\delta\left(\int{d^2}\xi\sqrt{g_c}
\chi\right)
\delta\left(\oint{d}{s_c}\phi\right){e^{-{S_1}
[\chi,\phi,{g_c}]}}.
\]

\noindent Here $\chi$ is the the quantum flutuation around the
classical solution and $\phi$ is the free value it takes on the
boundary. The metric ${g_c}^{ab}$ is given by
${e^{{\varphi_c}}}{\hat{g}^{ab}}$ and
the one loop action is

\[
{S_1}[\chi,\phi,{g_c}]={1\over{2\gamma}}\int{d^2}\xi\sqrt{g_c}
{{g_c}^{ab}}{\partial_a}\chi{\partial_b}\chi-{1\over{2\gamma}}{\eta_c}
\int{d^2}\xi\sqrt{g_c}{\chi^2}-{1\over{4\gamma}}{k_c}\oint{d}{s_c}
{\phi^2}.
\]

\noindent Let us separate $\chi$ into a fixed background field $\chi_b$
and an homogeneous Dirichlet field $\bar{\chi}$. Introducing the
operator ${\mathcal{O}_c}={\Delta_c}-{\eta_c}$ we specify $\chi_b$
as the solution to the boundary-value problem
${\mathcal{O}_c}{\chi_b}=0$, ${\chi_b}{|_B}=\phi$. Thus:

\begin{eqnarray*}
&\Gamma(A,L)={e^{-{S_L}[{\varphi_c},\hat{g},{p_c},{q_c}]}}\int
{\mathcal{D}_{g_c}}\phi\,\delta\left(\oint{d}{s_c}{\chi_b}\right)
\exp\left(-{1\over{2\gamma}}\oint{d}{s_c}{\chi_b}{\partial_{nc}}
{\chi_b}\right)\nonumber\\
&\int{\mathcal{D}_{g_c}}\bar{\chi}\,
\delta\left[\int\sqrt{g_c}\left(\bar{\chi}+{\chi_b}\right)\right]
\exp\left(-
{1\over{2\gamma}}\int{d^2}\xi\sqrt{g_c}\bar{\chi}{\mathcal{O}_c^D}
\bar{\chi}\right).
\end{eqnarray*}

\noindent We can use the delta function for the integral along the
boundary of $\phi$ to eliminate the constant zero mode of the covariant
Laplacian $\Delta_c$. However unlike the closed string case we still have
another delta function which involves the other orthogonal modes of
$\chi$. Unfortunately this means we are left with a functional integral
too difficult to be solved here.

All these calculations can be attempted taking homogeneous Neumann
boundary conditions on the Liouville field $\partial_{\hat n}\varphi=0$.
The results for the critical
exponents using the scaling argument are the same. However we run into
difficulties in performing the semi-classical expansion because the
classical solution $\varphi_c$
does not satisfy homogeneous Neumann boundary conditions so if the full
Liouville field does, then the classical field and the quantum
fluctuation are not independent, but rather are related with each other
on the boundary
$\partial_{\hat n}\varphi_c+\partial_{\hat n}\chi=0$. So we conclude
that the free boundary conditions are much better suited for the
semi-classical expansion.

\subsection{The tachyon gravitational scaling dimensions}

Let us now calculate the gravitational scaling
dimensions of the tachyon vertex operators for free boundary conditions.
For the anomalous gravitational scaling dimension of the bulk tachyon
vertex operator we consider the expectation value of the 1-point
function at fixed area $A$

\begin{eqnarray*}
&<{W_j}>(A)={1\over{\Gamma(A)}}\int{\mathcal{D}_{\hat{g}}}
(\bar{\phi},\bar{\Psi})d{\bar{\Psi}_0}
{{\left(\oint{d}\hat{s}\right)}^{1/2}}
{e^{-{S_c^{00}}[\bar{\Psi},{\bar{\Psi}_0},
\hat{g}]-{\bar{S}^0}[\bar{\phi},\hat{g}]}}\nonumber\\
&\delta\left(\int{d^2}\xi\sqrt{\hat{g}}{e^{\alpha\phi}}-A\right)
\int{d^2}{\xi_j}
\sqrt{\hat{g}}{e^{i{p_j}\cdot{X}({\xi_j})}}
{e^{{\gamma_j}\phi}}.
\end{eqnarray*}

\noindent By definition the bulk gravitational scaling dimension is
as in the closed string $<{W_j}>(A)\sim{A^{1-{\Delta_j}}}$. Applying
the scaling argument we find ${\Delta_j}=1-{\gamma_j}/\alpha$ and this
leads to the KPZ equation for the anomalous
gravitational dimension in the open string:

\[
{\Delta_j}-{\Delta_j^0}=-{\alpha^2}{\Delta_j}({\Delta_j}-1).
\]

Similarly we define the anomalous gravitational scaling dimension of the
boundary tachyon vertex operator by
$<{W_j^B}>(A)\sim{A^{1/2-{\Delta_j^B}}}$. Then the scaling argument
gives ${\Delta_j^B}={\Delta_j}/2$.

We can also define critical exponents associated with the expectation
values at fixed
length $L$. These should also be interpreted as anomalous gravitational
scaling dimensions. In this case the asymptotic limits are
$<{W_j}>(L)\sim{L^{1-{\Delta_j}}}$ and
$<{W_j^B}>(L)\sim{L^{1/2-{\Delta_j^B}}}$, where $\Delta_j$ and
$\Delta_j^B$ are given as in the case of fixed area $A$.

\subsection{A connection with matrix models}

These results generalise to other models and physical
systems. As we observed before the open string is a toy model for the
$c\leq1$ boundary conformal field theories \cite{JC} coupled to 2D
quantum gravity.
In the next section we show that similar results can be written down for
this more realistic class of models. Here we finish by considering a
comparison with exact results of matrix models at genus zero \cite{MSS}.
According to ref. \cite{MSS} we may deduce from matrix models
calculations the following exact
expression for $\Gamma(A,L)$ when the surface has the topology of a disc:

\[
\Gamma(A,L)={A^x}{L^y}{e^{-{L^2}/A}},
\]
where $x=-Q/\alpha$ and $y=-3+Q/\alpha$. This formula is consistent with
our scaling laws given in eqs. (\ref{42}), (\ref{50}) and (\ref{51}).
Introducing it in the
definitions of $\Sigma(A)$ and $\Omega(L)$ we find:

\[
\sigma(1)=x+y/2+7/2,\quad\omega(1)=2x+y+5.
\]
When we substitute back the values of $x$ and $y$ we get
the same results for $\sigma(1)$ and $\omega(1)$ as we did using the David,
Distler and Kawai's scaling argument.

This is an indication that our results should be in agreement with those
obtained in models of dynamically triangulated open random surfaces.
However it should be emphasised that a full comparison is
beyond the scope of the present work.

\section{Minimal Models On Open Random Surfaces}

The open string analysis can now be easily extended to
$c\leq1$ minimal conformal field theories on open random surfaces if we
represent the matter sector by a conformally extended Liouville theory.
The curious affinity between the matter and
gravitational sector Liouville theories that emerges
for closed surfaces generalises
to the case with boundaries.
We simply take the
matter action of eq. (\ref{30}) with additional boundary terms:

\begin{eqnarray*}
&{S_M}[\Phi,\tilde{g}]={1\over{8\pi}}\int{d^2}\xi\sqrt{\tilde{g}}
\left[
{1\over{2}}{\tilde{g}^{ab}}{\partial_a}\Phi{\partial_b}\Phi+i\left(
\beta-1/\beta\right)\tilde{R}\Phi\right]+\nonumber\\
&+{i\over{4\pi}}\left(
\beta-1/\beta\right)\oint{d}\tilde{s}{k_{\tilde{g}}}\Phi\
+{\mu^2}\int{d^2}\xi\sqrt{\tilde{g}}\left({e^{i\beta\Phi}}+
{e^{-i/\beta\Phi}}\right)+\nonumber\\
&+\lambda\oint{d}\tilde{s}\left[
{e^{i\beta\Phi/2}}+{e^{-i/(2\beta)\Phi}}\right].
\end{eqnarray*}

\noindent This is the conformally extended  Toda field theory defined
on an open surface for the Lie algebra $A_1$. It has recently been
considered as a Coulomb gas
description of the $c\leq1$ minimal conformal matter in the
case of Neumann boundary conditions imposed on the matter field
\cite{JS}. Here we assume without proof that the same is true of
when the matter satisfies free boundary
conditions. In fact for both free and Neumann boundary
conditions we
have a full Weyl invariant non-critical theory at the quantum level
to all orders in the Coulomb gas perturbation theory.

For
definiteness we take here the free boundary conditions on all fields.
The central charge of the matter theory is
${c_M}=1-6{{\left(\beta-1/\beta\right)}^{2}}$.
Requiring that the sum of this and the central charges of
the gravitational sector Liouville field and the reparametrisation
ghosts vanish gives $\gamma=\pm{i}\beta$, where $\gamma$ relates to our
previous string $Q$, $Q=i(\gamma+1/\gamma)$. The Liouville field
renormalisation parameter must satisfy the equation
$1-\alpha(\beta+1/\beta)+{\alpha^2}=0$ which, as before, gives us two
branches ${\alpha_+}=\beta$ and ${\alpha_-}=1/\beta$. All the boundary
renormalisation parameters relate to $\alpha$ and $\gamma$ as happened
for the string case. We find dressed vertex operators of
vanishing conformal weight on the bulk

\[
{U^D}(jj')=\int{d^2}\xi\sqrt{\hat{g}}\exp\left[\left(l\beta+
{{l'}\over{\beta}}\right)
\phi\right]\exp\left[-i\left(j\beta-{{j'}\over{\beta}}\right)\Phi\right]
\]

\noindent where $l=-j$, $l'=j'+1$ or $l=j+1$, $l'=-j'$. On the boundary
we also define dressed primary vertex operators of vanishing conformal
weight consistent with the need to consider the Liouville field as an
arbitrary Weyl scaling on the whole surface

\[
{U_B^D}(jj')=\oint{d}\hat{s}\exp\left[\left(l\beta+{{l'}\over{\beta}}
\right){\phi\over{2}}
\right]\exp\left[-i\left(j\beta-{{j'}\over{\beta}}\right){\Phi\over{2}}
\right].
\]

\noindent As occurred for the string, Dirichlet boundary
conditions on
the Liouville field
imply that we have no dynamical quantum degrees of freedom on the
boundary, and hence no boundary vertex operators. Although they
still allow the cancellation of the Weyl anomaly provided the metric
has a discontinuity as the boundary is approached.

The open string
formulas for the critical exponents generalise to these models. Thus the
susceptibility exponent is
$\sigma({\chi_o})=2-{\chi_o}Q/(2\alpha)$, the Feynman mass exponent is
$\omega({\chi_o})=2-{\chi_o}Q/
\alpha$. The
semi-classical limit is obtained for $\beta\to+\infty$ and, just like
for closed surfaces, selects the classical branch ${\alpha_+}=\beta$.
As in the open string the saddle point expansion singles out the free
boundary conditions on the Liouville field. Similarly
we find the same expressions for the anomalous gravitational scaling
dimensions of the primary vertex operators. In the end the
gravitational scaling dimension of a boundary operator is half that
of a bulk operator, the latter being related to its bare
conformal dimension by the KPZ equation.

\section{Conclusions}

In this paper we have shown how to extend the approach of David, Distler
and
Kawai to the coupling of boundary conformal field theories to 2D quantum
gravity. The organising principal behind their approach is Weyl
invariance at the quantum level applied to a perturbative expansion
analogous to the
Coulomb gas. We used this to determine the
renormalised parameters,
gravitational dressings and surface critical
exponents such as the susceptibility of random surfaces, the anomalous
gravitational scaling dimensions of primary vertex operators and the
Feynman mass exponent. The
crucial problem is the choice of boundary
conditions on the Liouville field. We have discussed free, Neumann and
Dirichlet boundary
conditions on the Liouville field. The first two lead to similar results
within this perturbative approach, but Dirichlet conditions imply that the
metric is discontinuous as the boundary is approached. We have also
considered the
semi-classical expansion and advocated the free boundary conditions
for the Liouville field, since homogeneous Neumann boundary conditions
do not allow a clean split between the classical and quantum pieces of
the field, but rather couple them together.
As would be expected the bulk properties are equal
for open and closed surfaces.
This approach may also be
naturally extended to higher genus and more complex boundary
structures.
Unfortunately as for closed surfaces the results only apply to the
weak coupling of $c\leq1$ boundary conformal field theories to gravity.
In the case of the open string this means unrealistic target space
dimensions $d\leq1$. Finally, we found the same close affinity
between the matter sector when represented by a Liouville theory and
the gravitational sector in this weak Coulomb gas
phase as occurs in the case of closed surfaces.
\vspace{1cm}

\centerline{\bf{Note added}}

\vspace{0.25cm}
After submitting this paper we were informed of refs. \cite{JAS} and
\cite{JAM}. In ref. \cite{JAS} the open string 2D quantum gravity with
Neumann boundary conditions has been analysed. The results agree with
ours. In ref. \cite{JAM} it is conjectured that Neumann and free boundary
conditions are equivalent, although as our discussion shows the free
boundary conditions are in fact better suited to the semi-classical
expansion.
\vspace{1cm}

\centerline{\bf{Appendix A}}

\vspace{0.25cm}

\leftline{THE NON-LOCAL WEYL CHANGE OF ${\hat{G}_K}(\xi,\xi')$}

\vspace{0.25cm}

Start by taking eq. (\ref{15}) and multiply it by $d\hat{s}(\xi)$. Since
$d\hat{s}(\xi)d\hat{s}(\xi''){\hat{K}_D}(\xi,\xi'')$ and
$d\hat{s}(\xi){\hat{\delta}_B}(\xi-\xi')$ are Weyl invariant use
${\delta_{\rho}}d\hat{s}(\xi)=(1/2)\rho(\xi)d\hat{s}(\xi)$ to get

\begin{equation}
d\hat{s}(\xi)\oint{d}\hat{s}(\xi''){\hat{K}_D}(\xi,\xi''){\delta_{\rho}}
{\hat{G}_K}(\xi'',\xi')=-{{\rho(\xi)d\hat{s}(\xi)}\over{2\oint{d}\hat{s}
(\eta)}}+
{{d\hat{s}(\xi)\oint{d}\hat{s}(\zeta)\rho(\zeta)}\over{2{{\left[\oint
{d}\hat{s}(\eta)\right]}^2}}}\label{57}.
\end{equation}
Next multiply eq. (\ref{57}) by ${\hat{G}_K}(\xi,\xi''')$ and integrate
on $\xi$. Using eqs. (\ref{15}) and (\ref{55}) find:

\[
{\delta_\rho}{\hat{G}_K}(\xi,\xi')=-
{{\oint{d}\hat{s}(\xi''){\hat{G}_K}(\xi'',\xi)\rho(\xi'')}\over{2\oint{d}
\hat{s}(\xi''')}}+{{\oint{d}\hat{s}(\xi''){\delta_\rho}{\hat{G}_K}
(\xi'',\xi')}\over{\oint{d}\hat{s}(\xi''')}}
\]
Finally the Weyl transformation of eq. (\ref{55})

\[
\oint{d}\hat{s}(\xi''){\delta_\rho}{\hat{G}_K}(\xi'',\xi')=
-{1\over{2}}\oint{d}\hat{s}(\xi'')\rho(\xi''){\hat{G}_K}(\xi'',\xi')
\]
leads to eq. (\ref{56}).
\vspace{1cm}

\centerline{\bf{Appendix B}}

\vspace{0.25cm}

\leftline{ANOMALY CANCELLATION FOR NEUMANN 2D QUANTUM GRAVITY}

\vspace{0.25cm}

The charge conservation selection rule is
$\int{d^2}\xi\sqrt{\hat{g}}{{\hat{J}}_N^{MM'}}=0$. Here we have written
${{\hat{J}}_N^{MM'}}=Q\hat{R}+{\bar{Q}_B}{k_{\hat{g}}}
{\hat{\delta}_B^2}
-8\pi\alpha{\sum_{P=1}^M}
{\delta^2}(\xi-{\xi_P})/\sqrt{\hat{g}(\xi)}-
4\pi{\alpha_B}{\sum_{P=1}^{M'}}
{\hat{\delta}_B^2}(\xi-{\xi_P})$, where $\int{d^2}\xi\sqrt{\hat{g}}
{\hat{\delta}_B^2}=\oint{d}\hat{s}$. The Neumann Green's
function satisfies

\[
\hat{\Delta}{\hat{G}_N}(\xi,\xi')={{{\delta^2}(\xi-\xi')}\over
{\sqrt{\hat{g}(\xi)}}}-{1\over{\int{d^2}\xi''\sqrt{\hat{g}(\xi'')}}},
\quad
{\partial_{\hat{n}}}{\hat{G}_N}(\xi,\xi')=0,
\]

\[
\int{d^2}\xi\sqrt{\hat{g}(\xi)}{\hat{G}_N}(\xi,\xi')=0.
\]
We find the non-local
functional:

\[
{\mathcal{F}_N^{MM'}}={1\over{16\pi}}\int{d^2}\xi'{d^2}\xi''
\sqrt{\hat{g}(\xi')}{\hat{J}_N^{MM'}}(\xi'){\hat{G}_N}(\xi',\xi'')
\sqrt{\hat{g}(\xi'')}{\hat{J}_N^{MM'}}(\xi'').
\]

Consider just one bulk Liouville vertex
operator. Then:

\begin{eqnarray*}
&{\delta_\rho}{\mathcal{F}_N^{10}}={Q\over{8\pi}}\int{d^2}\xi
\sqrt{\hat{g}}
{\hat{J}_N^{10}}\rho+{\alpha^2}\rho({\xi_1})-{Q\over{8\pi}}
\int{d^2}\xi
\sqrt{\hat{g}}{\hat{J}_N^{10}}{\delta_\rho}\ln\int{d^2}\xi\sqrt{\hat{g}}-
\nonumber\\
&-{1\over{8\pi\int{d^2}\xi\sqrt{\hat{g}(\xi)}}}\int{d^2}\xi
\sqrt{\hat{g}}{\hat{J}_N^{10}}\int{d^2}\xi'{d^2}\xi''
\sqrt{\hat{g}(\xi')}\rho(\xi'){\hat{G}_N}(\xi',\xi'')
\sqrt{\hat{g}(\xi'')}{\hat{J}_N^{10}}(\xi''),
\end{eqnarray*}
where ${\bar{Q}_B}=2Q$. This eliminates
the terms in
${\partial_{\hat{n}}}\rho$. The $\alpha^2$ term comes from the
Weyl change of ${\hat{G}_N}(\xi,\xi)$, $\xi\not\in{B}$. We have:

\[
{\hat{G}_N}(\xi,\xi')=\hat{G}(\xi,\xi')+{\hat{H}_N}(\xi,\xi')
\]
where ${\hat{H}_N}(\xi,\xi')$ is defined by:

\[
\hat{\Delta}{\hat{H}_N}(\xi,\xi')=-{1\over{\int{d^2}\xi''
\sqrt{\hat{g}(\xi'')}}},\quad{\partial_{\hat{n}}}{\hat{H}_N}
(\xi,\xi')=-{\partial_{\hat{n}}}\hat{G}(\xi,\xi').
\]
Then the local change is
${\delta_{\rho}}{\hat{G}_{N\varepsilon}}(\xi,\xi)=\rho(\xi)/(4\pi)$.
Thus $Q=\pm\sqrt{(25-d)/6}$ and $1-\alpha{Q}+{\alpha^2}=0$.

Now consider just one boundary Liouville vertex
operator. When $\xi=\xi'$ is on the boundary, ${\hat{H}_N}(\xi,\xi)$ is
divergent because $\hat{G}(\xi,\xi)$ is singular. In a
neighbourhood of order $\sqrt{\varepsilon}$ around $\xi$ the shape of
the boundary is flat. Then ${\hat{G}_N}(\xi,\xi)$ is defined by the
method of
images so that ${\hat{H}_N}(\xi,\xi)=\hat{G}(\xi,\xi)$. Hence the local
change now is
${\delta_{\rho}}{\hat{G}_{N\varepsilon}}(\xi,\xi)=\rho(\xi)/(2\pi)$.
Thus ${\alpha_B}=\alpha$.

\end{document}